\documentclass[traditabstract]{aa}  

\usepackage{graphicx}
\usepackage{rotating}
\usepackage{natbib}
\usepackage{txfonts}
\usepackage{threeparttable}
\usepackage{float}

\newcommand{\co}[0]{CoRoT}

\begin{document} 

\title{Spectral characterization and differential rotation study \\
       of active CoRoT stars}

\subtitle{}

\author{
E. Nagel\inst{1}
\and
S. Czesla\inst{1}
\and
J. H. M. M. Schmitt\inst{1}
}
   
\institute{
Hamburger Sternwarte,
Gojenbergsweg 112, 21029 Hamburg\\
\email{evangelos.nagel@hs.uni-hamburg.de}
}

\date{Received date; Accepted date}

\abstract
{The CoRoT space telescope observed nearly 160\,000 light curves. Among 
the most
outstanding is that of the young, active planet host star CoRoT-2A. In addition to
deep planetary transits, the light curve of CoRoT-2A shows strong rotational
variability and a superimposed beating pattern.
To study the stars that produce such an intriguing pattern of photometric
variability, we identified a sample of eight stars
with rotation periods between 0.8 and 11 days and photometric variability
amplitudes of up to 7.5\,\%,
showing a similar CoRoT light curve. We also
obtained high-resolution follow-up spectroscopy with TNG/SARG and
carried out a spectral analysis with SME and MOOG.
We find that the color dependence of the light curves
is consistent with rotational modulation due to starspots and that
latitudinal differential rotation provides a viable
explanation for the light curves, although starspot evolution is also
expected to play an important role.
Our MOOG and SME spectral analyses provide consistent results,
showing that the targets are dwarf stars with
spectral types between F and mid-K. Detectable \ion{Li}{i} absorption in 
four
of the targets confirms a low age of $100-400$\,Myr also deduced from
gyrochronology. Our study indicates that the photometric beating phenomenon
is likely attributable to differential rotation in fast-rotating stars 
with outer
convection zones.}

\maketitle

\section{Introduction}
Dark surface spots are among the most prominent manifestations of solar and 
stellar activity. While they typically cover less than one percent of the solar 
surface, their coverage fractions on highly active stars can reach tens
of percent \citep[e.g.,][]{ONeal1996}. Therefore, the surfaces of active stars
are not always homogeneously bright.
Starspots are ultimately caused by the stellar magnetic field, which is sustained
by a stellar dynamo. Although mostly spatially unresolved,
the photometric and spectral variability induced by
rotating stellar surfaces can be used to study the photospheres of active stars 
using techniques such as Doppler imaging and light curve inversion 
\citep[e.g.,][]{Rodono1986, Piskunov1990, Karoff2013}.

While long-term ground-based 
photometric observation campaigns have long been used to study stellar
activity \citep[e.g.,][]{Jarvinen2005, Olah2006}, the recent advent of the space-based
observatories CoRoT and \textit{Kepler} \citep{Baglin2006, Jenkins2010} offers
photometric data of unprecedented temporal cadence, continuity, and accuracy.
Although the main objectives of both missions are the search for extrasolar
planets and asteroseismological studies, the data are also extremely interesting
in the context of stellar activity.

CoRoT-2A is among the most active planet host-stars known to date.
The star shows strong \ion{Ca}{ii} H and K emission line cores and X-ray
emission as well as \ion{Li}{i} absorption \citep{Bouchy2008}, 
suggesting an age of $\sim300$\,Myr \citep{Schroeter2011}.
The broad band light curve of the CoRoT-2 system is among the most
remarkable discoveries made by the CoRoT mission.
In addition to deep transits caused by
a bloated hot Jupiter \citep{Guillot2011}, the light curve
shows photometric variability on at least two distinct timescales
\citep[e.g.,][]{Alonso2008, Lanza2009, Czesla2009, Huber2010}.
First, there is clear evidence for rotational variability with a period of
$\sim 4.5$ days. Second, the amplitude of the rotation-induced pattern is
itself variable, showing modulation reminiscent of a ``beating pattern'' with a period
of roughly 50 days. The general pattern remained stable for at least 140\,d, i.e.,
the duration of the CoRoT observation. In their analyses of the spot
configuration, \citet{Lanza2009} and \citet{Huber2010} reconstructed two active
longitudes on opposing hemispheres. The prominent photometric beating pattern is 
related to an alternation in the strength of these active longitudes, 
in combination with differential rotation.

A similar photometric behavior, however, at a much longer
timescale, has been observed in the active late-type giant
\mbox{\object{FK Comae Berenices}} \citep[e.g.,][]{Jetsu1993}. Here, the pattern has also
been attributed to a pair of opposing active longitudes of alternating
strengths. A change in the center of activity, i.e., starspot coverage fraction,
to the opposing hemisphere is known as a ``flip-flop'' event \citep{Jetsu1993,
Olah2006, Hackman2013}.
The discovery of such flip-flops in \mbox{\object{FK Com}} was later supplemented
with the observation of flip-flops in nearly a dozen additional objects such as
\mbox{RS CVn}-type stars \citep[e.g.,][]{Berdyugina1998} and young solar analogs
\citep{Berdyugina2005}. These observations have inspired a number of
theoretical works on the magnetic field configurations capable of reproducing 
the observed behavior \citep[for an overview see][]{Berdyugina2006}.

\begin{table*}
\caption{Parameters of the observed stars provided by the \textit{Exo-Dat} information system.}		
\label{table:1}      
\centering		
\begin{tabular}{ccccccccc} 
\hline\hline               
CoRoT-ID & Right Ascension & Declination & B & V & R & I & Spectral & Luminosity  \\
         &    (J2000.0)    &  (J2000.0)  &   &   &   &   & type     & class       \\
\hline
102577568 & 06\,40\,48.3 & $-01\,05\,15.5$ & $12.60$ & $11.81$ & $11.56$ & $11.12$ & G8 & V  \\
102601465 & 06\,41\,28.7 & $+01\,03\,31.3$ & $14.09$ & $13.31$ & $12.93$ & $12.51$ & G8 & V  \\
102606401 & 06\,41\,35.3 & $-01\,27\,26.1$ & $12.66$ & $12.03$ & $11.82$ & $11.37$ & F0 & V  \\
102656730 & 06\,42\,44.8 & $-00\,28\,24.6$ & $13.65$ & $12.91$ & $12.60$ & $12.26$ & G5 & V  \\
102743567 & 06\,44\,37.9 & $-00\,44\,13.4$ & $12.38$ & $11.99$ & $11.85$ & $11.62$ & F0 & V  \\
102763571 & 06\,45\,04.9 & $+00\,59\,09.3$ & $13.94$ & $13.08$ & $12.69$ & $12.21$ & K2 & V  \\
102778303 & 06\,45\,24.7 & $-00\,22\,41.3$ & $13.78$ & $12.64$ & $12.07$ & $11.45$ & K1 & III\\
102791435 & 06\,45\,42.4 & $-00\,17\,39.0$ & $13.80$ & $12.84$ & $12.39$ & $11.94$ & K0 & V  \\
\hline
\end{tabular}
\end{table*} 

In this paper, we present the analysis of the CoRoT photometry and follow-up
high-resolution spectra of a sample of eight CoRoT targets. The sample stars 
were selected by their photometric variability which show a variation pattern
similar to that of CoRoT-2A. 
Our main observational objective is to find common spectral characteristics and
to establish a link between the structure of the light curves and the stellar 
parameters. In particular, we are interested in the nature of the observed 
activity pattern and whether this pattern is powered by differential rotation. 
We also provide estimates of stellar age and distance 
in an attempt to thoroughly characterize a larger 
number of stars with a well-observed beating behavior to significantly 
increase the data base on which a theory of the underlying dynamo processes can be established.
 
Our paper is organized as follows. In Sect.~\ref{section:target} we describe
the sample selection and the photometric/spectroscopic data. The analysis of the
data is presented in Sects.~\ref{section:analysis} and
Sect.~\ref{section:spectral}. We discuss our findings in
Sect.~\ref{section:discussion} and, finally, present our conclusions in
Sect.~\ref{section:conclusions}.

\section{Observations}
\label{section:target}

\subsection{CoRoT photometry and target selection}
CoRoT was a space-based 30\,cm telescope dedicated to stellar photometry
\citep{Auvergne2009}. The satellite observed a few thousand stars simultaneously
with a temporal cadence of either 32\,s or 512\,s. The CoRoT mission was divided
into several runs, i.e., continuous observing periods pointed at a
specific section of the sky.
CoRoT provided simultaneous three-color photometry in a red, a green, and
a blue channel \citep{Auvergne2009}. Although the exact
passbands of these channels remain essentially unknown and are expected to vary 
between individual targets and observations, they can still be used to test the
plausibility of physical assumptions.
The sum of the signals registered in the individual color channels is known as
the ``white'' light curve. 

We checked the light curves of about 11\,000 stars situated in CoRoT's first
two Long Run fields (LRc01 and LRa01) for variability and manually grouped the
light curves into variability classes. In the course of this analysis, we discovered
about 40 objects with photometric characteristics similar to those of CoRoT-2A.
The light curves of these objects show a dominant modulation between 2
and 11 days, which is most likely caused by rotation. Additionally, there is a
beating pattern found with $P_{\mathrm{beat}} \gg P_{\mathrm{rot}}$. On the
rotational and beating timescale, the photometric variability is typically a
few percent. 

Eight of the brightest targets were selected for spectroscopic follow-up
observations. These stars are listed in Table~\ref{table:1}, along with their
available observational stellar parameters from the \textit{Exo-Dat}
database\footnote{The values in Table~\ref{table:1} are rounded. The
exact numbers including uncertainties can be found at
\url{http://cesam.oamp.fr/exodat/}} \citep{Deleuil2009}. 
This database contains photometric information from several 
catalog references. To provide uniform photometry, we give B, V, R, and I 
magnitudes from the \mbox{OBSCAT} catalog. The spectral type and luminosity class were 
derived using spectral energy distribution (SED) analysis \citep{Deleuil2009}.
However, such assigned spectral types and luminosity
classes have to be used with some caution.

\begin{table}[!t]
\caption{CoRoT light curves of sample targets.
\label{tab:CoRoTObs}}
\begin{tabular}{c c c c c} \hline\hline
CoRoT-ID & LRa01 & Sampling & LRa06 & Sampling \\
   &       & [s]      &       & [s] \\ \hline
102577568 & \checkmark & 512 &  &  \\
102601465 & \checkmark & 512/32\tablefootmark{a} & &  \\
102606401 & \checkmark & 32 & &  \\
102656730 & \checkmark & 512/32\tablefootmark{a} & \checkmark & 32 \\
102743567 & \checkmark & 512 & \checkmark & 32 \\
102763571 & \checkmark & 512 & &  \\
102778303 & \checkmark & 512 & \checkmark & 32 \\
102791435 & \checkmark & 512 & \checkmark & 32 \\
\hline
\end{tabular}
\tablefoot{\tablefoottext{a}{Sampling changed during run}}
\end{table}

The light curves of our sample stars were observed in the frame of
the first Long Run (LRa01), which lasted for about
131\,d starting in October 2007. Later, a subsample of four
targets was reobserved during sixth Long Run (LRa06) with a duration of
77\,d, which began in January 2012; the observations
are summarized in Table~\ref{tab:CoRoTObs}.
We used the N2-level pipeline data available from the
CoRoT Data Center\footnote{\url{http://idoc-corot.ias.u-psud.fr/}}.

\begin{figure*}
\begin{center}
\includegraphics[width=0.49\textwidth]{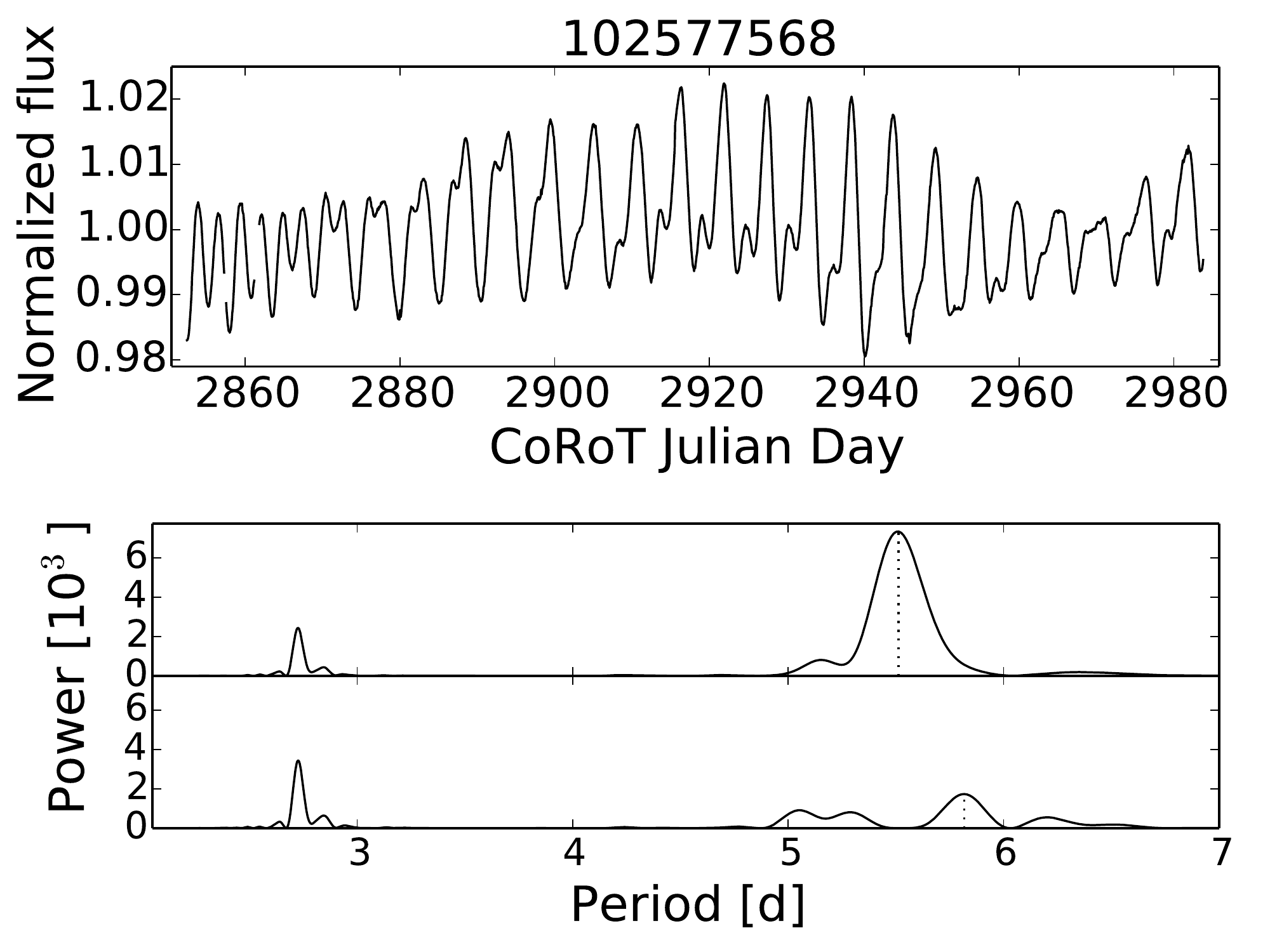}
\includegraphics[width=0.49\textwidth]{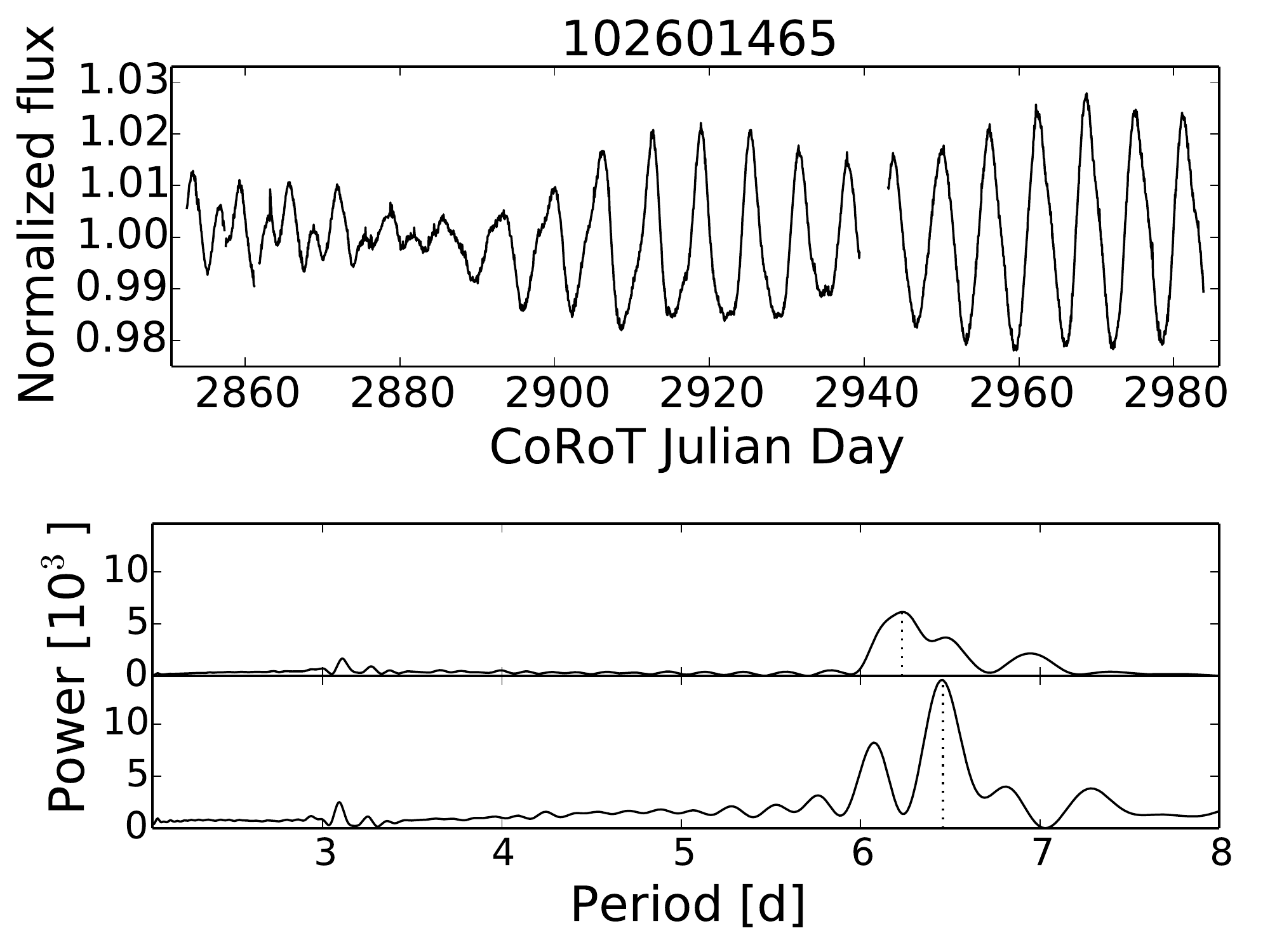}
\includegraphics[width=0.49\textwidth]{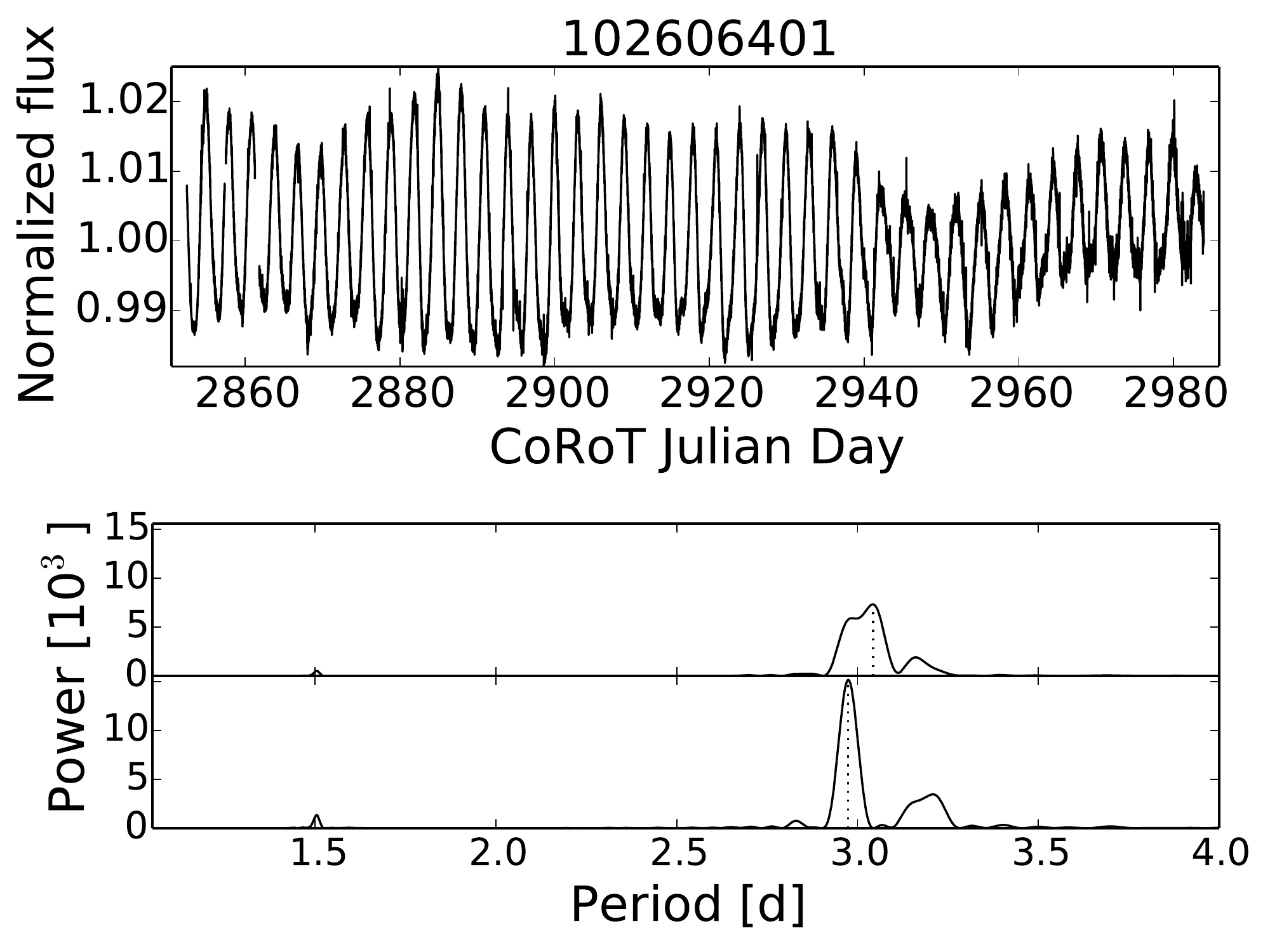}
\includegraphics[width=0.49\textwidth]{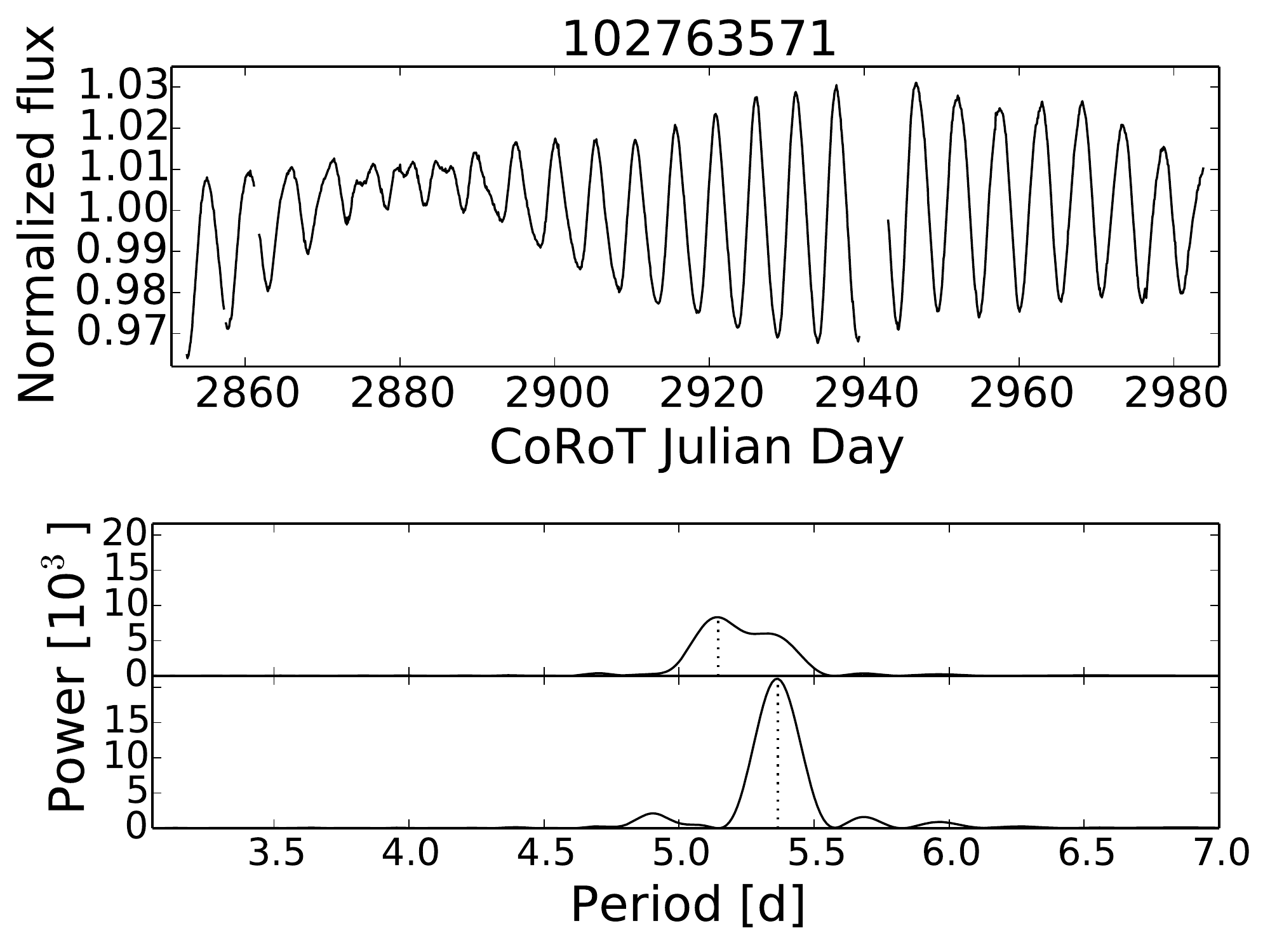}
\caption{\label{figure:lc1} Light curves and periodograms of \object{CoRoT 102577568}, \object{CoRoT 102601465}, \object{CoRoT 102606401}, and \object{CoRoT 102763571} (LRa01). The origin of the CoRoT Julian day is 1 January 2000 12:00.00.}
\end{center}
\end{figure*}

\begin{figure*}
\begin{center}
\includegraphics[width=0.49\textwidth]{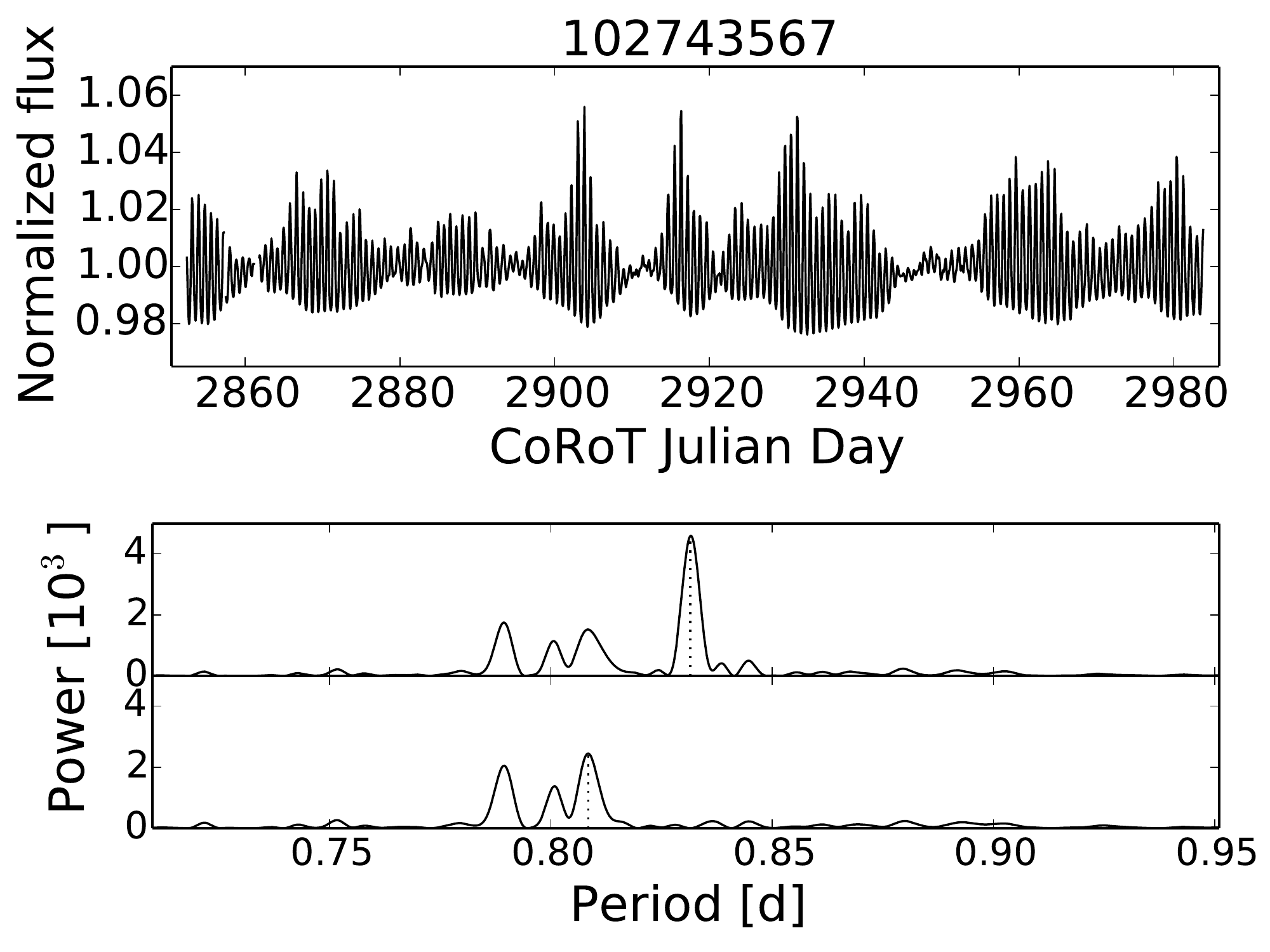}
\includegraphics[width=0.49\textwidth]{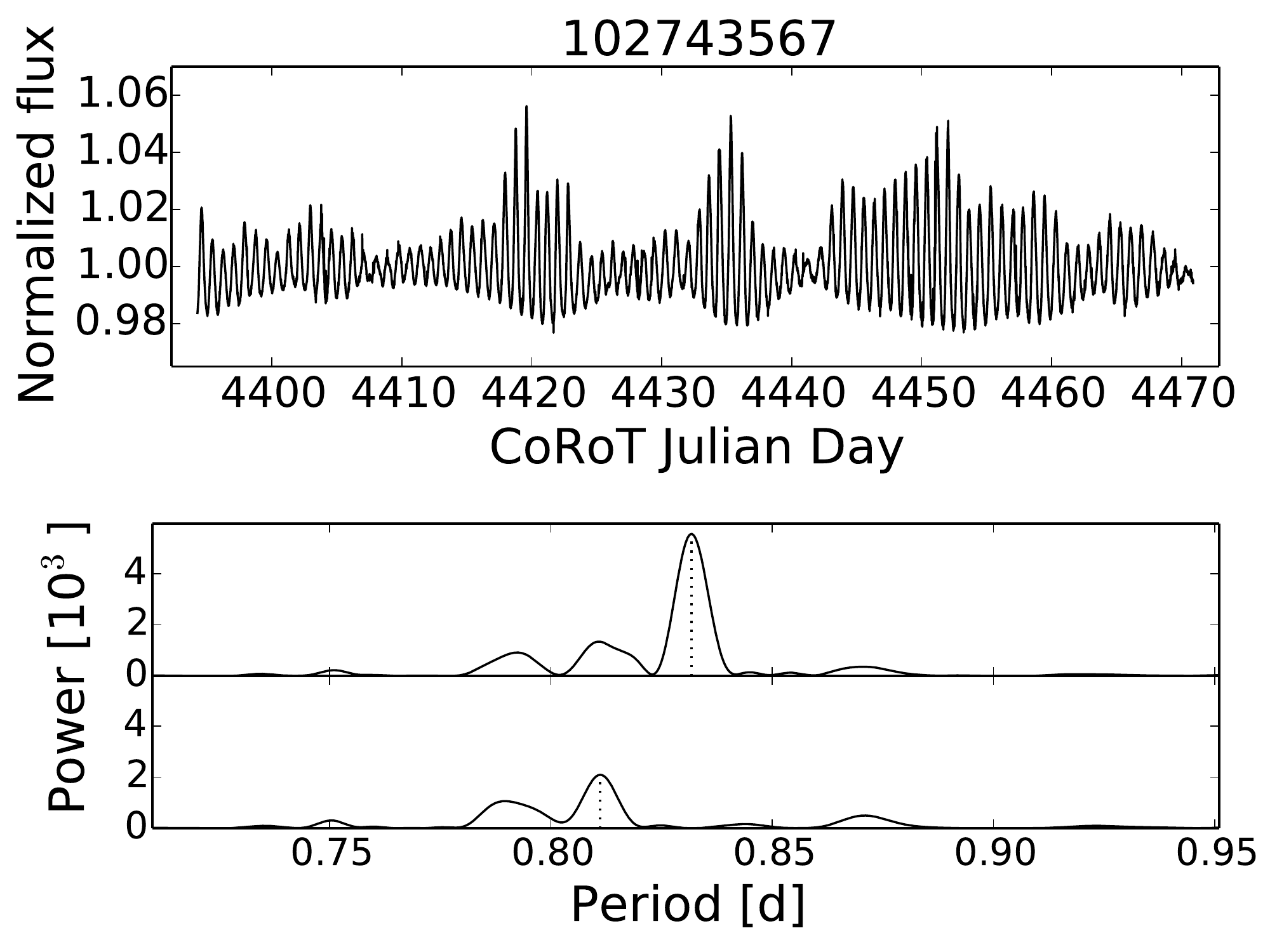}
\includegraphics[width=0.49\textwidth]{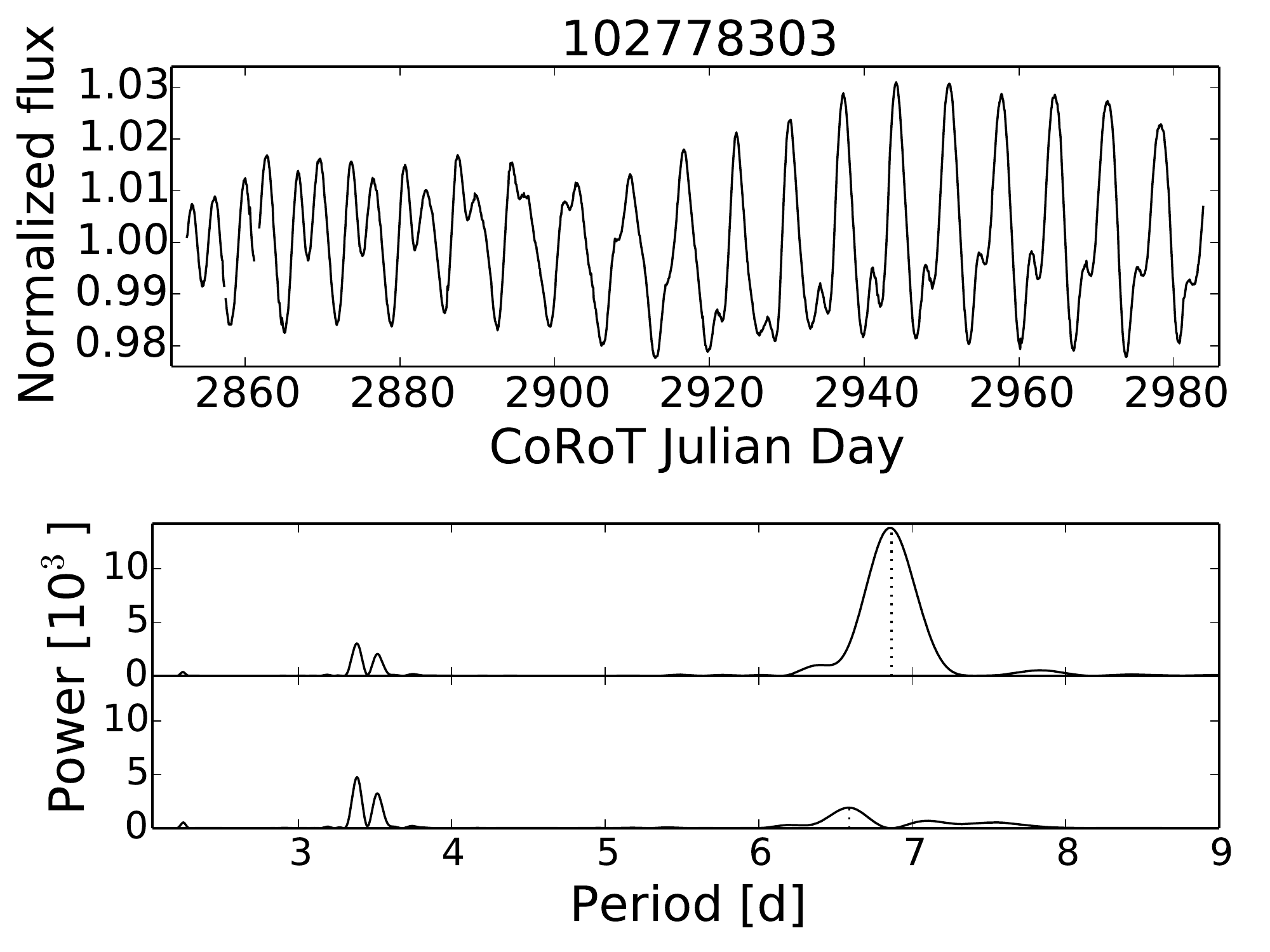}
\includegraphics[width=0.49\textwidth]{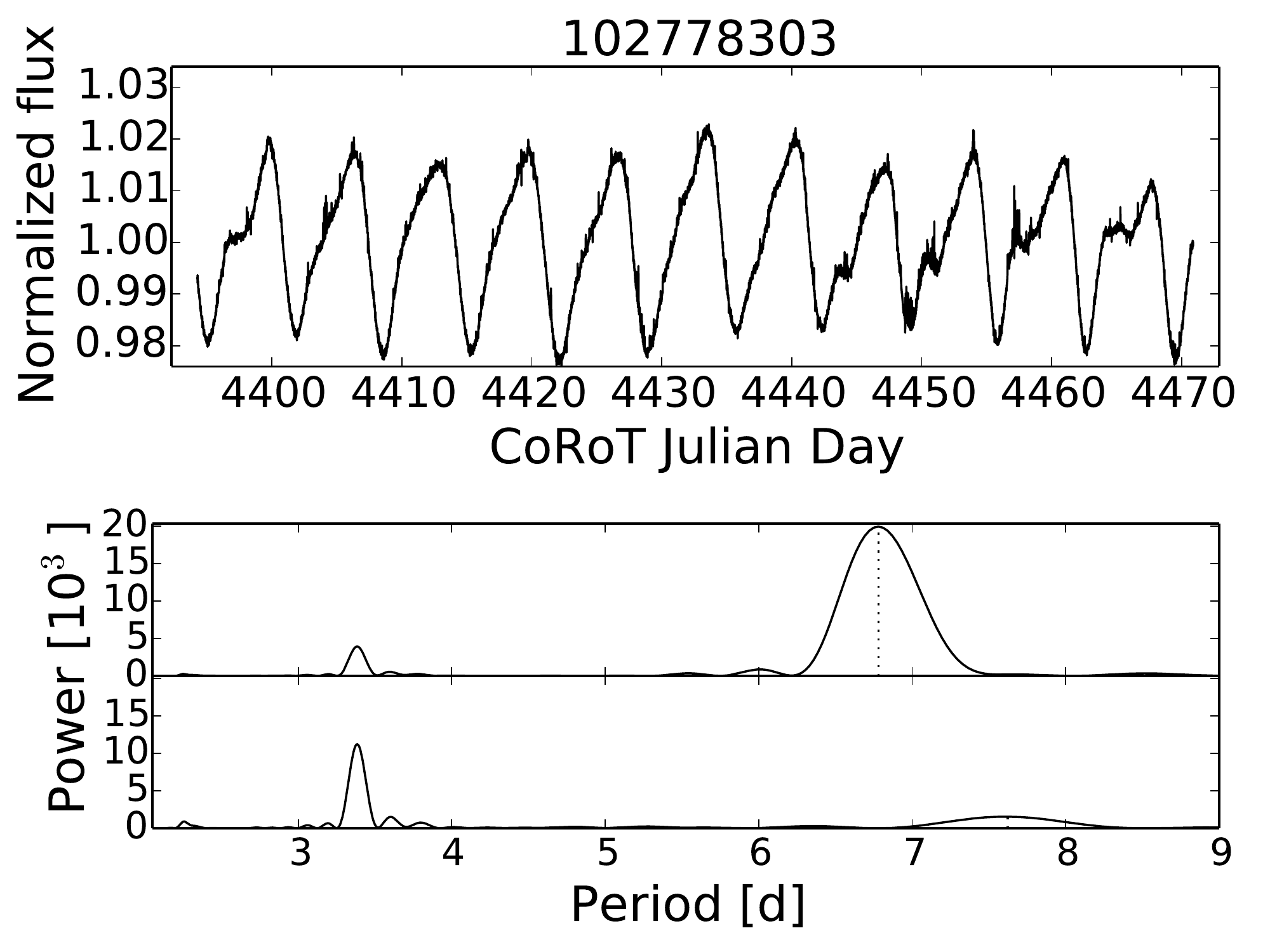}
\caption{\label{figure:lc2} Light curves and periodograms of \object{CoRoT 102743567} and \object{CoRoT 102778303}
(left: LRa01, right: LRa06).}
\end{center}
\end{figure*}

\begin{figure*}
\begin{center}
\includegraphics[width=0.49\textwidth]{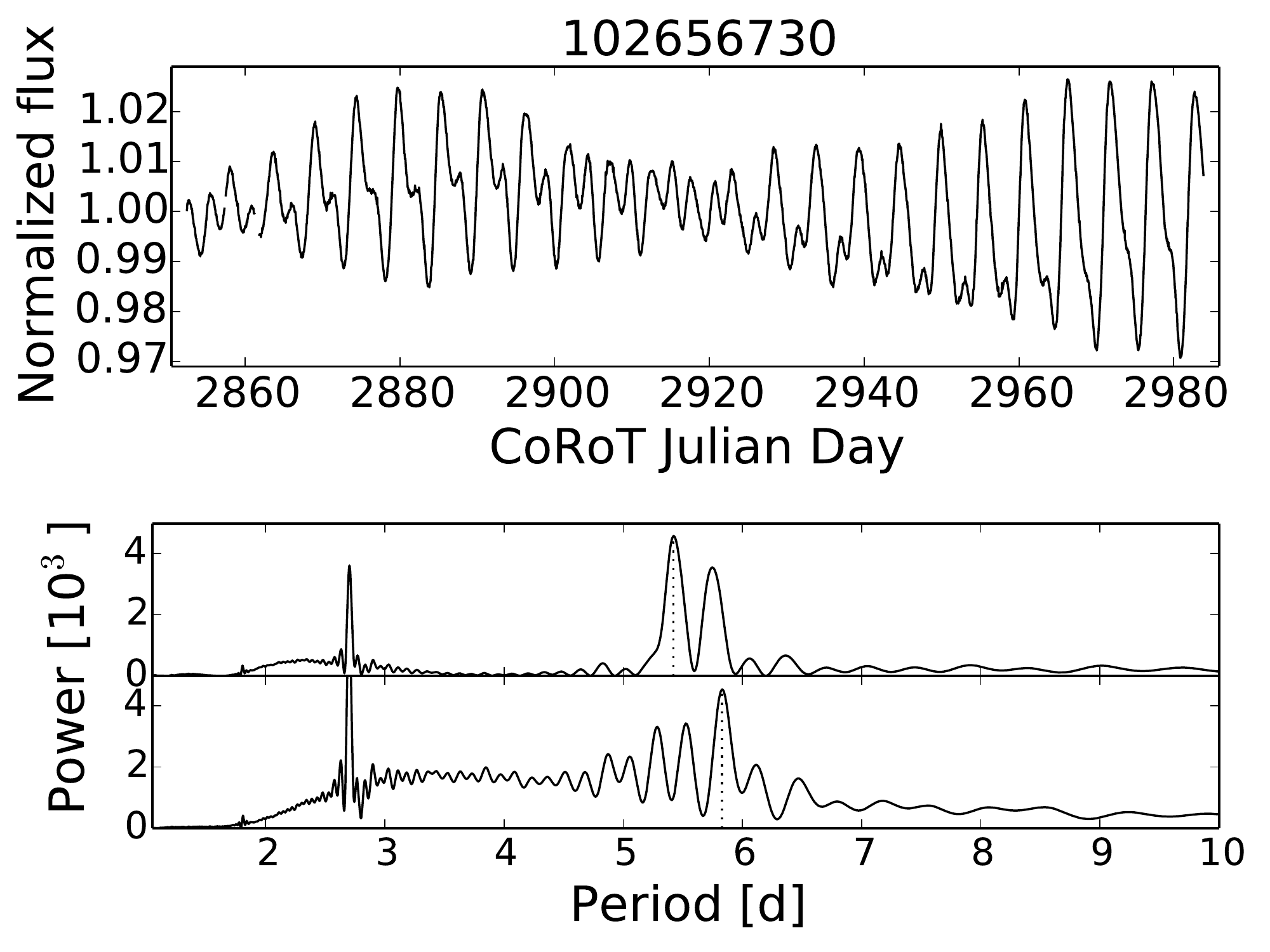}
\includegraphics[width=0.49\textwidth]{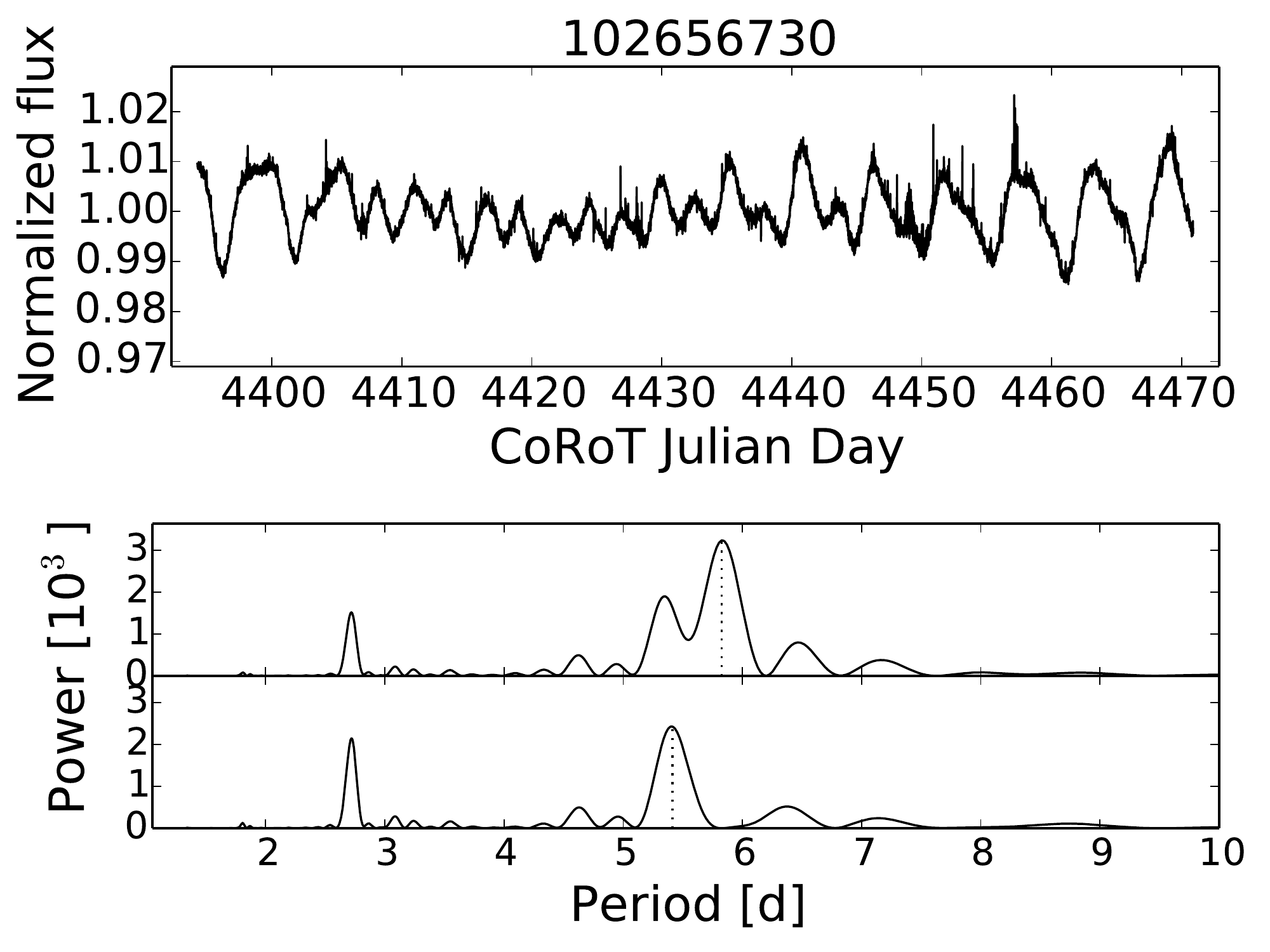}
\includegraphics[width=0.49\textwidth]{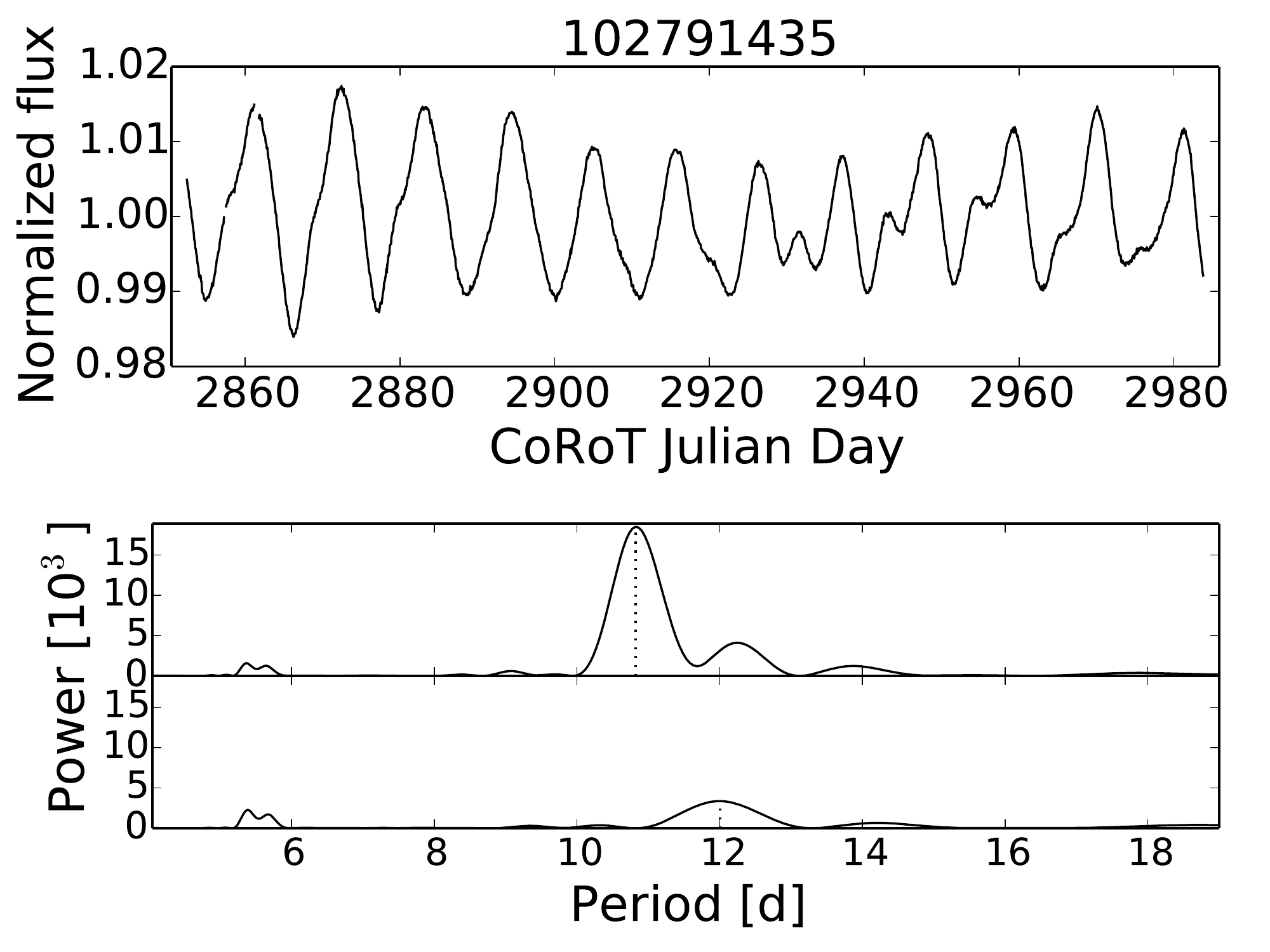}
\includegraphics[width=0.49\textwidth]{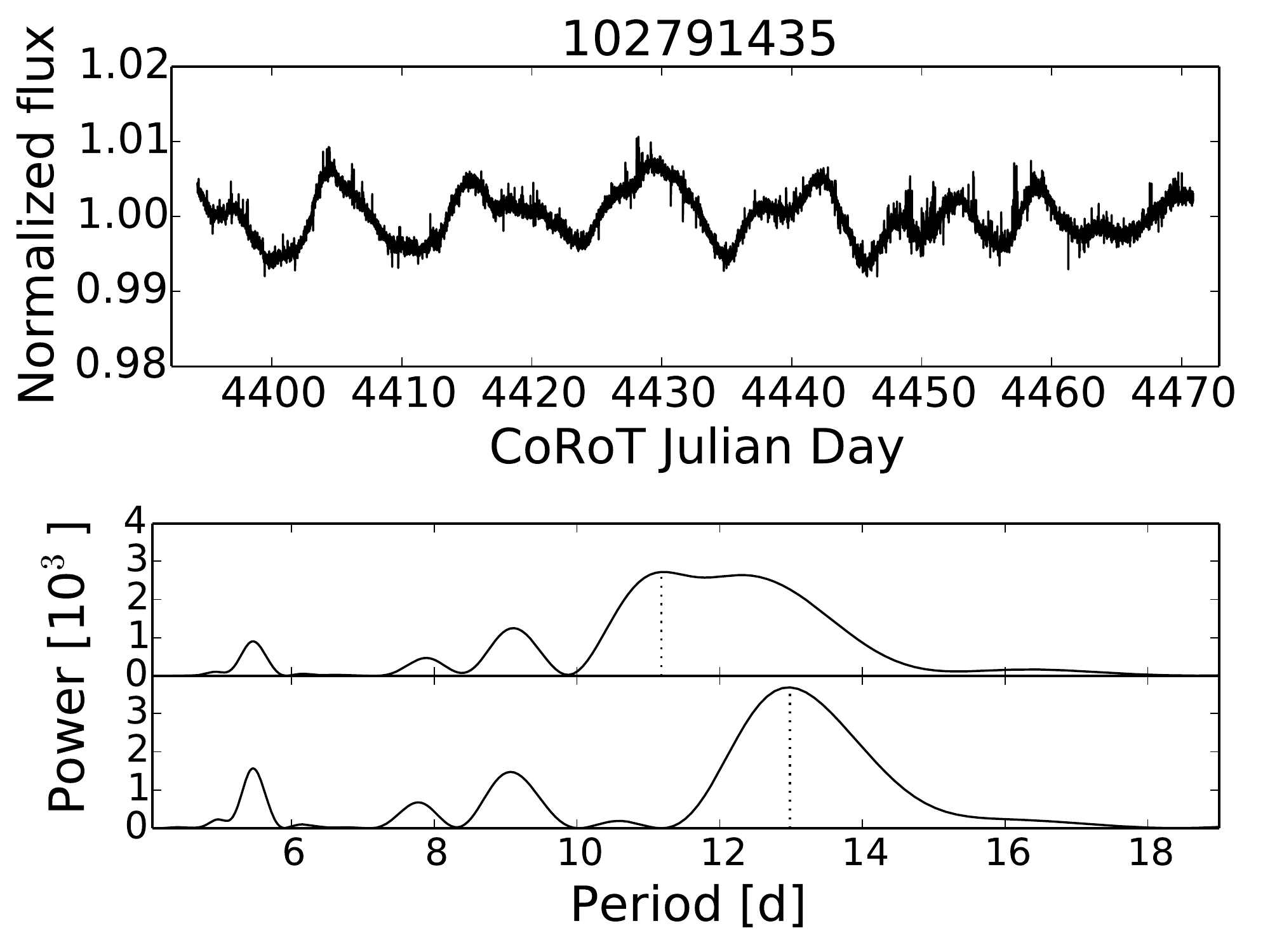}
\caption{\label{figure:lc3} Light curves and periodograms of \object{CoRoT 102656730} and \object{CoRoT 102791435}
(left: LRa01, right: LRa06).}
\end{center}
\end{figure*}

After acquiring the data we first
removed all data points marked as ``bad quality'' by the CoRoT
pipeline. Such points are produced, for example, by the impact of high-energy
particles during CoRoT passages of the South
Atlantic Anomaly. Second, we normalized the light curves by dividing by the
mean flux. In a final step, we detrended the light curves
to remove instrumental long-term trends by subtracting a linear model.
The resulting white light curves are shown in the upper
panels of \mbox{Figs.~\ref{figure:lc1} -- \ref{figure:lc3}}. For the purpose of
visualization, we rebinned the light curves except that of CoRoT 102743567,
which shows a particularly fast pattern of variability.
In particular, we binned by a factor of 8 for light curves
with a 512\,s sampling rate and by a factor of 16 for those with a 32\,s sampling. 

For all our light curves
we estimated the photometric error, $\sigma$. To this end, we divided the
light curves into consecutive 0.5\,d long chunks, fitted them using a
second-order polynomial, and determined the standard deviation of the
residuals. The mean standard deviation determined in the chunks was used as
an error estimate on individual data points. The photometric error was used as 
an input
parameter when we utilized the generalized Lomb-Scargle periodogram, 
which takes the photometric error of the light curve into account 
(see Sect.~\ref{subsection_period}).

\subsection{High-resolution spectroscopy}
The spectroscopic observations of our target stars were carried out between 
8 and 12 December, 2011, with the SARG echelle spectrograph, 
mounted on the 3.58\,m ``Telescopio Nazionale Galileo'' (TNG) on La Palma. We
used the yellow cross-disperser, which provides wavelength coverage in the
$5200-7800\,\AA$ range with a gap of $150\,\AA$ at around $6200\,\AA$
caused by the separation of the CCD detectors.
Our setup yields a spectral resolution of $R\sim 57\,000$.

During our observations, weather conditions were unstable with
periods of excellent conditions interrupted by rather cloudy phases. Therefore,
seeing varied between $0.6\,''$ and $5\,''$. 
As part of our observational campaign, we also obtained a solar spectrum by
observing the light reflected from the asteroid 15\,Eunomia,
which is required for a differential abundance analysis.

The data reduction was carried out using the
REDUCE package developed by \citet{Piskunov2002}. In particular, we performed a
bias correction, order definition, extraction of the
blaze function, and flat fielding. The treatment of scattered light and hot pixels 
is described by \citet{Piskunov2002}.
The wavelength calibration is based on
ThAr-lamp reference frames and was carried out using the WAVECAL extension of
REDUCE. Finally, a barycentric velocity correction was applied.

The majority of the targets was observed several times with individual 
exposure times between 1800\,s and 3600\,s.
To improve the signal-to-noise ratio (S/N), we averaged all spectra taken 
for an individual target. For every spectrum we then computed a S/N in the
$6578-6580\,\AA$ interval, which contains no strong spectral features.
The resulting values are listed in Table~\ref{table:snr} along with the total
integration time. Here, the S/N refers to the averaged spectrum
and although it varies across individual
echelle orders and across the entire spectrum, these numbers
broadly characterize the data quality. Finally, we manually continuum-normalized 
the spectra by dividing the flux by linear functions, which were adjusted to 
have the same gradient as the continuum.

\begin{table}
\caption{Spectroscopic data obtained with SARG.}
\label{table:snr}
\centering
\begin{tabular}{ccc}  
\hline\hline          
CoRoT-ID & S/N & Total exposure time\\
         &     & [s]         \\  
\hline
102577568 & 43 & 1800 \\
102601465 & 37 & 3600 \\
102606401 & 69 & 1800 \\
102656730 & 55 & 12600\\
102743567 & 35 & 9900 \\
102763571 & 44 & 1800 \\
102778303 & 46 & 1800 \\
102791435 & 60 & 2700 \\
\hline
\end{tabular}
\end{table}

\section{CoRoT light curves analysis} 
\label{section:analysis}

In the upper panels of Figs.~\ref{figure:lc1} -- \ref{figure:lc3}, we show the
white-band \co\ light curves of our target stars. All light curves show
pronounced variability with peak-to-peak amplitudes of several percent (see
Table~\ref{table:period}), which we attribute to a rotating and temporally
evolving starspot pattern.

\subsection{Starspot induced color changes}
To assess the plausibility of the starspot hypothesis to explain the observed
variability, we examined the individual \co\ color channels and their relation.
As observed on the Sun, we postulate that the temperature of the starspots
is lower than the remaining stellar photosphere
\citep[e.g.,][]{Strassmeier2009}.
Once a spot appears on the visible hemisphere, the flux in all three color
channels decreases. However, because the spot is cooler, the stellar spectrum
appears redder. Therefore, the decrease in the blue channel flux is expected to
be stronger than in the red channel. The opposite is the case, when a spot
rotates off the visible hemisphere. 

\begin{figure}
\begin{center}
\includegraphics[width=0.5\textwidth]{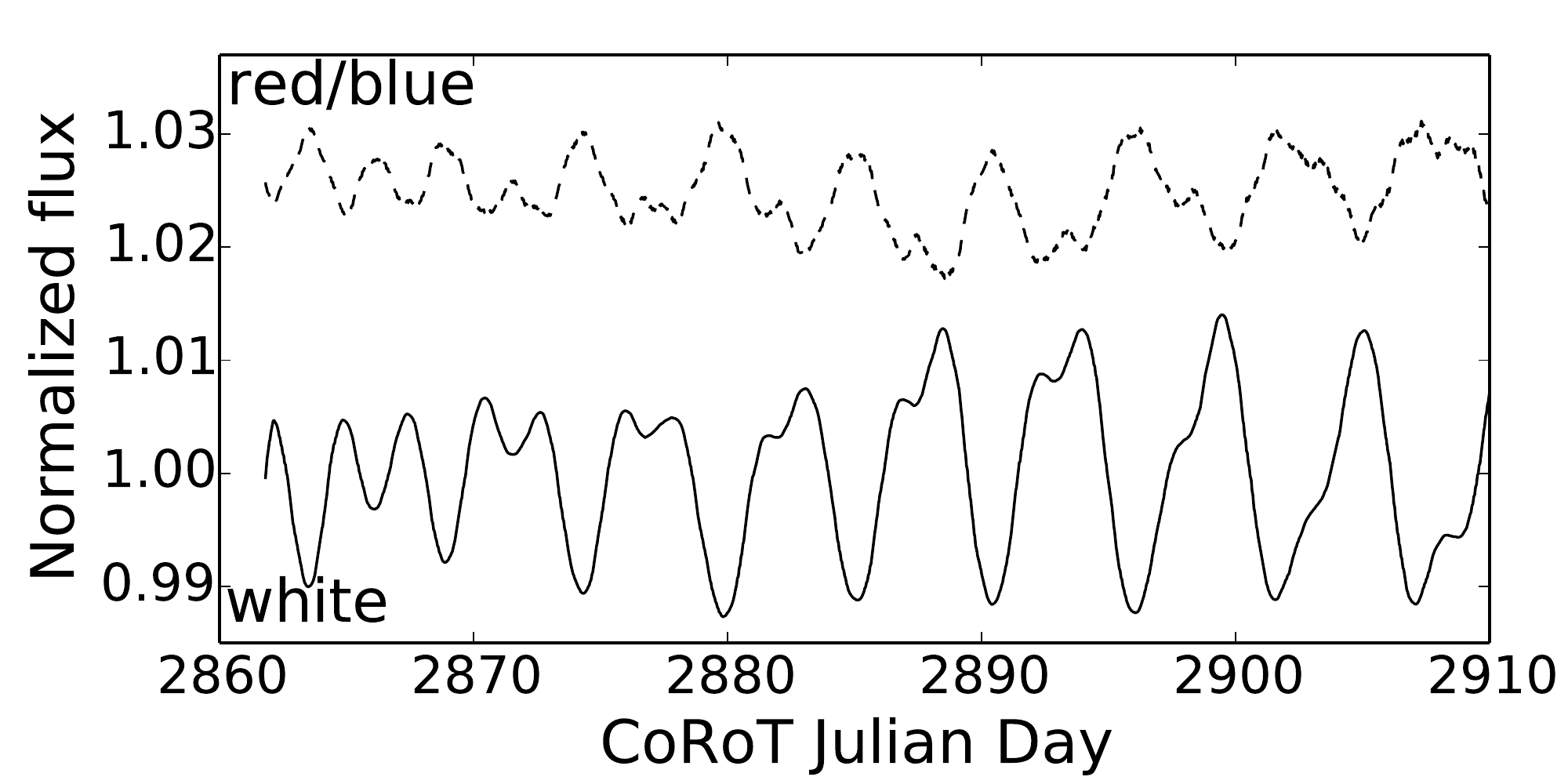}
\caption{\label{figure:spot} Excerpt of the light curve of CoRoT 102577568 obtained in
    the white band (solid line) along with the ratio of the red- and
    blue-channel light curves (dashed, shifted upward by 0.025 for clarity).}
\end{center}
\end{figure}

In Fig.~\ref{figure:spot} we show 
an excerpt of the white-band light curve of CoRoT 102577568 along with
the ratio of the fluxes recorded in the
red and blue channels. Clearly, both curves are anticorrelated. A decrease
in the white light flux is accompanied by a reddening of the star, which
is consistent with spot-induced photometric modulation.
We verified that all investigated light curves show this behavior. 
Alternatively, also pulsations could cause a similar signature.
Yet, typical photometric pulsation amplitudes
on solar-like stars fall far behind the observed modulation
\citep[$<10^{-5}$,][]{White2011}.
Therefore, we conclude that the modulation in the
light curves is indeed dominated by starspots with only a marginal
contribution from pulsations.

\subsection{Comparing LRa01 and LRa06}
\label{sec:LRa01VsLRa06}

For half of our targets, two light curves observed about 4 years apart are
available (see Table~\ref{tab:CoRoTObs}). A visual comparison of the two light
curves allows to assess the temporal stability of the variability pattern.
For CoRoT 102743567 and CoRoT 102778303 both the amplitude 
and appearance of the light curves remained
unchanged. For CoRoT 102656730 this may also be the case, although it seems less clear.
During the second, shorter observation, the amplitude is smaller and the pattern
appears somewhat more chaotic. However, this is compatible with an observation in
the low-amplitude phase of the beating pattern clearly visible during the first
Long Run. The situation is similar for CoRoT 102791435, whose light curve appears more
erratic, but still periodically variable during the later Long Run. 

\subsection{Period analysis}
\label{subsection_period}

Clearly, the light variation of our target stars shows some periodic components. 
To study the variability in the frequency domain, we applied the generalized 
Lomb-Scargle periodogram \citep{Zechmeister2009} to all light curves and show the
results in the upper subfigures in the lower panels of 
\mbox{Figs.~\ref{figure:lc1} -- \ref{figure:lc3}}.

All periodograms show distinguished peaks at periods between about one and
twelve days. We attribute these peaks and the associated modulation in
the light curve to rotating starspots and, therefore, also identify
the associated period with the stellar rotation period.
Generally, periodograms obtained from stellar light curves observed in LRa01 and their
counterparts observed in LRa06 show similar structures.
Peaks are generally less well resolved in the LRa06 periodograms, because the
observation is only about half as long as the LRa01 data set.

\begin{table*}
\caption{Measured peak-to-peak variability amplitude, $A_{\mathrm{pp}}$, and fitted
rotation periods, $P_{\mathrm{fit,1}}$ and $P_{\mathrm{fit,2}}$. Calculated beating period 
as well as absolute and relative horizontal shear derived from $P_{\mathrm{fit,1}}$
and $P_{\mathrm{fit,2}}$.}
\label{table:period}     
\centering              
\begin{tabular}{ccrr|rcc}
\hline\hline\\[-3.5mm]                
CoRoT-ID &  $A_{\mathrm{pp}}$ [\%] & \multicolumn{1}{r}{$P_{\mathrm{fit,1}}$
[d]} & \multicolumn{1}{r|}{$P_{\mathrm{fit,2}}$ [d]} & 
\multicolumn{1}{r}{$P_{\mathrm{beat}}$} [d] &
$\Delta\Omega_{\mathrm{beat}}$ [rad\,d$^{-1}$] & $\alpha$\\\\[-3.5mm]
\hline\\[-3mm]
\multicolumn{7}{c}{2007} \\
\hline
 102577568 & 3.8 &  5.51$\pm$0.11 &  5.82$\pm$0.09 & 103.4 & 0.061 & 0.053 \\
 102601465 & 4.6 &  6.23$\pm$0.13 &  6.45$\pm$0.10 & 182.7 & 0.034 & 0.034 \\
 102606401 & 3.7 &  3.04$\pm$0.04 &  2.97$\pm$0.03 & 129.0 & 0.049 & 0.023 \\
 102656730 & 5.2 &  5.43$\pm$0.07 &  5.83$\pm$0.07 & 79.1 & 0.079 & 0.069 \\
 102743567 & 7.5 &  0.832$\pm$0.002 & 0.809$\pm$0.002 & 29.3 & 0.215 & 0.028 \\
 102763571 & 5.9 &  5.14$\pm$0.10 & 5.36$\pm$0.08 & 125.2 & 0.05 & 0.041 \\
 102778303 & 4.8 &  6.86$\pm$0.16 &  6.59$\pm$0.12 & 167.4 & 0.038 & 0.039 \\
 102791435 & 3.2 & 10.8$\pm$0.4 & 12.0$\pm$0.6 & 108.0 & 0.058 & 0.100 \\
\hline\\[-3mm]
\multicolumn{7}{c}{2012} \\
\hline
 102656730 & 2.7 &  5.83$\pm$0.15 & 5.41$\pm$0.13 & 75.1 & 0.084 & 0.072 \\
 102743567 & 7.0 &  0.832$\pm$0.003 & 0.811$\pm$0.004 & 32.1 & 0.196 & 0.025 \\
 102778303 & 4.2 &  6.78$\pm$0.25 & 7.61$\pm$0.38 & 62.2 & 0.101 & 0.109 \\
 102791435 & 1.2 & 11.2$\pm$0.9 & 13.0$\pm$0.9 & 80.9 & 0.078 & 0.138 \\
\hline                                                 
\end{tabular}
\end{table*}

The maximum-power periodogram peaks, which we call the primary peaks,
clearly show a tendency to be accompanied by at least one nearby secondary peak. A
particularly clear example for such a double-peak structure is the periodogram of
CoRoT 102656730 in Fig.~\ref{figure:lc3}. We adopted the 
prewhitening procedure described by \citet{Reinhold2013} to extract 
closely spaced periods from the periodogram. The period associated 
with the primary peak was used as input for a sine, whose amplitude 
and phase were fitted to the data. 
After subtracting the resulting sine from the light curve, we computed
the periodogram of the residuals. Following \citet{Reinhold2013}, we examined the period 
space within 30\,\% of the primary period, mainly to avoid alias periods. 
The resulting periodograms are shown in the lower subfigures in 
the lower panels of \mbox{Figs.~\ref{figure:lc1} -- \ref{figure:lc3}}.

To determine the periods associated with 
the main and secondary peaks, we fitted the peaks using a
Gaussian profile and used the center as an estimate of the peak location.
Additionally, we interpret their FWHM as an error estimate on the location.
The resulting periods and errors are listed in Table~\ref{table:period}, and
the locations of the primary and secondary peak are indicated by black vertical
dots in Figs.~\ref{figure:lc1} -- \ref{figure:lc3}.

In two out of the four stars observed twice by CoRoT we found that
the primary and secondary peaks in the LRa06 periodograms remained at
essentially the same location (CoRoT 102743567 and CoRoT 102656730). 
Another case is CoRoT 102778303, where the secondary
peak was already quite weak in the LRa01 periodogram. The structure is mostly 
washed out in the LRa06 periodogram. The situation is also different for 
CoRoT 102791435, where the secondary peak changed in position and power,
possibly indicating considerable change in the stellar spot configuration.

In the case of CoRoT 102601465, CoRoT 102606401, CoRoT 102763571 (all LRa01),
and CoRoT 102791435 (LRa06) the power of the secondary peak is larger 
than that of the primary peak. We were able to reproduce this behavior 
in simulated light curves generated by the superposition of sines with a continuum
of periods.
Nevertheless, we interpret the secondary peak and the associated period in
terms of differential rotation, but advise some caution in interpreting the
result.

We note that the light curves of CoRoT 102606401 and CoRoT 102601465 
were also included in the study
presented by \citet{Affer2012}, who provide period measurements of almost two
thousand stars observed by \co. For CoRoT 102606401 and CoRoT 102601465, 
they found rotational
periods of 3.039 and 6.375 days, which are compatible with our results.

\subsubsection{Differential rotation}
The presence of two similar periods in the
periodograms of stellar light curves has been attributed to differentially
rotating starspots \citep[e.g.,][]{Reinhold2013}. 
Differential rotation is also a possible explanation for the marked
beating pattern in the light curve of \object{CoRoT-2A} \citep[e.g.,][]{Lanza2009,
Huber2010} and, therefore, also of our target stars.

On the Sun the observed latitude-dependent rotation velocity can be 
parametrized by
\begin{equation}
  \Omega (\Psi) = \Omega_{\mathrm{eq}} - \Delta \Omega \, \sin^2 \Psi \; 
  = \Omega_{\mathrm{eq}} \left(1 - \alpha \sin^2\Psi \right) \; ,
  \label{eq:DR}
\end{equation}
where $\Psi$ denotes the latitude, $\Omega_{\mathrm{eq}}$ is the equatorial
angular velocity, $\Delta \Omega = \Omega_{\mathrm{eq}} -
\Omega_{\mathrm{pole}}$ is the ``absolute horizontal shear'', and $\alpha$ is
dubbed the ``relative horizontal shear'' \citep[e.g.,][]{Snodgrass1983, Reinhold2013}.
Similar to the situation on the Sun we may expect that the rotation rates of starspots depend not only on latitude but also on their anchoring depth.
Therefore, we caution that there may not be a unique relation between latitude and
starspot rotation rate as suggested by Eq.~\ref{eq:DR} and
helioseismology reveals that the rotation rate of the Sun depends on both
latitude and depth \citep{Schou1998}.

Therefore, if there are two measurements of $\Omega$ at two latitudes $\Psi_1$
and $\Psi_2$, $\Delta\Omega$ can be written as
\begin{equation}
  \Delta\Omega = \frac{\Omega(\Psi_1) - \Omega(\Psi_2)}{\sin^2(\Psi_2) -
  \sin^2(\Psi_1)} \; .
\end{equation}
Remaining ignorant of $\Psi_1$ and $\Psi_2$, as is often the case, a lower limit
on $|\Delta\Omega|$  may be obtained by assigning $\Psi_1=0^{\circ}$ and
$\Psi_2=90^{\circ}$, which is of course a rather unsatisfactory choice. 
Furthermore, attributing the larger rotational velocity measurement to the equator
($\Psi=0^{\circ}$), we ensure $\Delta\Omega>0$ and thus, solar-like differential
rotation.
As the angular velocity is related to the rotation period $P$, via 
$\Omega = 2\pi P^{-1}$, a lower limit on the absolute and relative
latitudinal shear may be obtained by
\begin{equation}
\Delta\Omega \geqslant 2\pi\left(\frac{1}{P_1} - \frac{1}{P_2} \right)
\;\;\;\mbox{and}  \;\;\; \alpha \geqslant \frac{P_2 - P_1}{P_2} \; ,
\label{eq:domega_alpha}
\end{equation}
where we demand $P_2 > P_1$ to endure solar-like differential rotation.

\citet{Froehlich2009} used a three-spot model to reproduce the light curve of
CoRoT-2A. Their modeling showed one slowly rotating and two fast rotating spot
components.
In particular, they found the length of the beating
period to be compatible with the overtaking period
\begin{equation}
\label{eq:P_beat}
P_{\mathrm{over}} = P_{\mathrm{beat}} = \left( \frac{1}{P_1} -
\frac{1}{P_2}\right)^{-1} \; 
\end{equation}
of the slowest and the fastest differentially rotating spots.
Tentatively assigning the primary and secondary peaks identified in the
periodograms to differentially rotating spots, we used Eq.~\ref{eq:domega_alpha}
to derive lower limits for the absolute and relative horizontal shear and
Eq.~\ref{eq:P_beat} to estimate the beating period; the values are listed in
Table~\ref{table:period}.

In our data, the beating period is not always well defined. Some
light curves do not fully cover a pattern as is the case for CoRoT 102601465
(Fig.~\ref{figure:lc1}) and the definition of the start and end of
a cycle remains somewhat ambiguous.
Nonetheless, the calculated beating periods are in reasonable agreement
with the observed behavior in the light curves.  
In CoRoT 102743567, a number of beating cycles with different duration was observed and the
beating structure persisted over (or reappeared after) four years. If
attributable to differential rotation, this indicates varying relative spot
velocities. Given a temporally constant and unique mapping between latitude and
velocity, this implies changing spot latitudes.

\section{Spectral analysis}
\label{section:spectral}

We determined the stellar parameters using two complementary approaches. First,
the curve-of-growth based MOOG package \citep{Sneden1973} and, second, the 
``Spectroscopy Made Easy'' (SME) package presented by \citet{Valenti1996}, 
which relies on direct spectral modeling. A visual inspection of the
spectra with regard to double line profiles did not reveal indications for 
binarity in any of our target stars.

\subsection{Stellar parameter determination using MOOG}
The stellar atmospheric parameters effective
temperature ($T_{\mathrm{eff}}$), surface gravity ($\log g$), metallicity 
([Fe/H]), and microturbulence velocity ($\xi_{\mathrm{mic}}$) were derived using
the 2014 version of MOOG\footnote{\url{http://www.as.utexas.edu/~chris/moog.html}}
and a grid of 1D Kurucz ATLAS9 model atmospheres\footnote{The Kurucz grids of
model atmospheres can be found at \url{http://kurucz.harvard.edu/grids.html}}
\citep{Kurucz1993}.
Additionally, we used PYSPEC, a Python interface to run MOOG \citep{Bubar2010}.

The analysis carried out by MOOG relies on a
curve-of-growth approach in which the stellar atmospheric equilibrium is
adjusted to match equivalent width (EW) measurements (see also \citet{Gray2005}).
To adjust the equilibrium and find the stellar parameters, it is essential to
measure the EWs of lines originating from a single element in
different ionization states. For low-mass
stars, iron provides a plethora of \ion{Fe}{i} and \ion{Fe}{ii}
lines distributed over the optical spectral range; the \ion{Fe}{ii} lines, however,
are usually more scarce.
Our choice of \ion{Fe}{i} and \ion{Fe}{ii} lines is based on the final line
list\footnote{The line list is available online at
\url{http://www.astro.up.pt/~sousasag/ares}} provided by
\citet{Sousa2008}, who present a total of 263
$\ion{Fe}{i}$ and 36 $\ion{Fe}{ii}$ lines, together with their excitation potentials
and oscillator strengths in the wavelength
range between $4500\,\AA$ and $6900\,\AA$.
We scrutinized their line list 
and eliminated lines not usable in our analysis, for example, lines suffering from
heavy blending or lines for which we could not derive adequate
continua to obtain the EW. Additionally, we neglected
weak lines ($\lesssim 5$\,m$\AA$) and lines in wavelength intervals with low S/N.
The number of the remaining $\ion{Fe}{i}$ and $\ion{Fe}{ii}$ lines of 
each star are listed in Table~\ref{tab:specan}. 

Equivalent widths were measured by fitting the line
profiles using a Gaussian. The local continuum was adjusted
manually for each line to achieve the optimal normalization.
To this end, a solar model spectrum served as a comparison, which we also used
for the unambiguous line identification in the observed spectra.
For CoRoT 102743567 no EWs could be determined from our spectra owing to extreme
rotational broadening ($\approx 64\,$km\,s$^{-1}$, see 
Sect.~\ref{Stellar_parameter_SME} and Table \ref{tab:specan}).

To minimize the impact of uncertainties in the atomic line parameters and the
characteristics of our particular instrumental setup on the analysis, we carried
out a differential abundance analysis.
Abundances are thus defined with respect to the Sun
\begin{equation} 
[\mathrm{Fe/H}] = \mathrm{log}
\left(\frac{N(\mathrm{Fe})}{N(\mathrm{H})}\right)_* - \mathrm{log} \left(
\frac{N(\mathrm{Fe})}{N(\mathrm{H})} \right)_{\odot},\quad \log\,
N(\mathrm{H}) \equiv 12.
\end{equation}
Therefore, we determined the EWs of our sample of spectral lines in a solar
spectrum, which we obtained by observing light reflected off the asteroid
15\,Eunomia with the same instrumental setup.
Our results are summarized in Table~\ref{tab:specan}. 

In MOOG, errors were estimated by studying various correlations. In
particular, the temperature value was varied until the correlation between
excitation potential and the $\ion{Fe}{i}$ abundances reached the $1\sigma$
level. The error on the microturbulence velocity was derived equivalently by
adapting it until the correlation between the
abundances and the reduced EWs\footnote{The EW divided by the
line wavelength} reached $1\sigma$. In MOOG,
the surface gravity is determined by minimizing the difference between the
abundances obtained from the $\ion{Fe}{i}$ and $\ion{Fe}{ii}$ lines.
Yet, the abundances depend on the stellar parameters including the surface
gravity, which implies that the error on the surface gravity depends on the
surface gravity itself. Therefore, the error had to be derived in an iterative
process, which is described in detail by \citet{Bubar2010}.

\begin{table*}
\caption{Results of spectral analysis with MOOG and SME.}
\label{tab:specan}      
\centering              
\begin{tabular}{clcrcrrc}        
\hline\hline              
CoRoT-ID & \multicolumn{1}{c}{$T_{\mathrm{eff}}$ [K]} & log\,$g$ &
\multicolumn{1}{c}{[Fe/H]\tablefootmark{a}}
& $\xi_{\mathrm{mic}}$ [$\mathrm{km\,s^{-1}}$] &
\multicolumn{1}{c}{$\varv_{\mathrm{rad}}$ [$\mathrm{km\,s^{-1}}$]} &
\multicolumn{1}{c}{$\varv\,\mathrm{sin}\,i$ [$\mathrm{km\,s^{-1}}$]} &
$N(\ion{Fe}{i}, \ion{Fe}{ii})$\tablefootmark{b}\\ \hline
\multicolumn{8}{c}{MOOG analysis} \\   
102577568 & $5600\pm110$ & $4.37\pm0.25$ & $-0.19\pm0.07$ & $1.94\pm0.13$ & \multicolumn{1}{c}{--} &
\multicolumn{1}{c}{--} & 104, 11\\
102601465 & $5630\pm110$ & $4.52\pm0.37$ & $0.01\pm0.07$ & $1.68\pm0.14$ & \multicolumn{1}{c}{--} &
\multicolumn{1}{c}{--} & 115, 11  \\
102606401 & $6270\pm140$ & $4.81\pm0.28$ & $0.02\pm0.09$ & $2.68\pm0.21$ & \multicolumn{1}{c}{--} &
\multicolumn{1}{c}{--} &79, 11   \\ 
102656730 & $5880\pm80$ & $4.84\pm0.19$ & $-0.13\pm0.05$ & $1.80\pm0.13$ & \multicolumn{1}{c}{--} &
\multicolumn{1}{c}{--} &119, 12  \\
102743567 & \multicolumn{1}{c}{--} & -- & \multicolumn{1}{c}{--} & -- & \multicolumn{1}{c}{--} & 
\multicolumn{1}{c}{--} & \multicolumn{1}{c}{--} \\         
102763571 & $5270\pm70$ & $4.24\pm0.17$ & $-0.22\pm0.04$ & $1.84\pm0.10$ & \multicolumn{1}{c}{--} &
\multicolumn{1}{c}{--} &137, 14  \\ 
102778303 & $4840\pm100$ & $4.09\pm0.52$ & $-0.54\pm0.05$ & $2.20\pm0.17$ & \multicolumn{1}{c}{--} &
\multicolumn{1}{c}{--} &119, 12  \\
102791435 & $5150\pm60$ & $4.39\pm0.26$ & $-0.03\pm0.03$ & $1.44\pm0.10$ & \multicolumn{1}{c}{--} &
\multicolumn{1}{c}{--} &135, 9  \\
\hline                
\multicolumn{8}{c}{SME analysis} \\ 
102577568 & $5690\pm90$ & $4.64\pm0.30$ & $0.06\pm0.07$ & $1.77\pm0.53$ &
$0.45\pm0.51$ & $8.7\pm1.4$ & \multicolumn{1}{c}{--} \\
102601465 & $5700\pm90$ & $4.88\pm0.31$ & $0.25\pm0.09$ & $1.77\pm0.42$ &
$4.1\pm0.4$ & $7.4\pm0.9$& \multicolumn{1}{c}{--} \\
102606401 & $6220\pm130$ & $4.93\pm0.32$ & $0.15\pm0.09$ & $1.76\pm0.41$ &
$-18.9\pm0.5$ & $13.2\pm1.1$& \multicolumn{1}{c}{--} \\
102656730 & $5900\pm100$ & $5.04\pm0.41$ & $0.02\pm0.10$ & $1.50\pm0.32$ &
$-16.9\pm0.5$ & $10.2\pm1.0$& \multicolumn{1}{c}{--} \\
102743567 & \multicolumn{1}{c}{--} & -- & \multicolumn{1}{c}{--} & -- &
$-46.9\pm0.2$ & $63.9\pm3.4$& \multicolumn{1}{c}{--} \\
102763571 & $5280\pm60$ & $4.61\pm0.25$ & $-0.15\pm0.09$ & $1.79\pm0.43$ &
$-12.1\pm0.4$ & $5.5\pm0.8$& \multicolumn{1}{c}{--} \\
102778303 & $4680\pm150$ & $4.97\pm0.19$ & $-0.23\pm0.30$ & $1.75\pm0.61$ &
$-12.3\pm0.5$ & $5.4\pm1.6$& \multicolumn{1}{c}{--} \\
102791435 & $5100\pm140$ & $4.69\pm0.25$ & $0.09\pm0.14$ & $1.54\pm0.46$  &
$-23.2\pm0.3$ & $3.6\pm0.8$& \multicolumn{1}{c}{--} \\
\hline
\end{tabular}
\tablefoot{\tablefoottext{a}{For MOOG this is specifically the iron
abundance. For SME, it refers to the metallicity pattern according to \citet{Grevesse2007}.}
\tablefoottext{b}{Number of \ion{Fe}{i} and \ion{Fe}{ii} lines used in the
MOOG analysis.}}
\end{table*}

\subsection{Stellar parameter determination using SME}
\label{Stellar_parameter_SME}

Following the analysis with MOOG,
we used the software package SME in version~2.1 to derive a complementary set of parameters.
In contrast to MOOG, the technique employed by SME is based
on fitting synthetic spectra to the observations. As input, a grid of
model atmospheres and an atomic line list are required.
For the latter we obtained atomic data from VALD3\footnote{VALD3 is
available at \url{http://vald.astro.uu.se/}} \citep{Kupka2000} and,
again, used ATLAS9 Kurucz atmospheres.

In our analysis, we used SME to fit individual echelle
orders neglecting the low S/N edges of the orders and other unsuitable
spectral regions. The latter comprise sections with spectral lines missing in
the line list; regions affected by cosmics, telluric lines, airglow emission
lines, and bands lacking any notable stellar absorption lines.

\begin{figure}
\begin{center}
\includegraphics[width=0.49\textwidth]{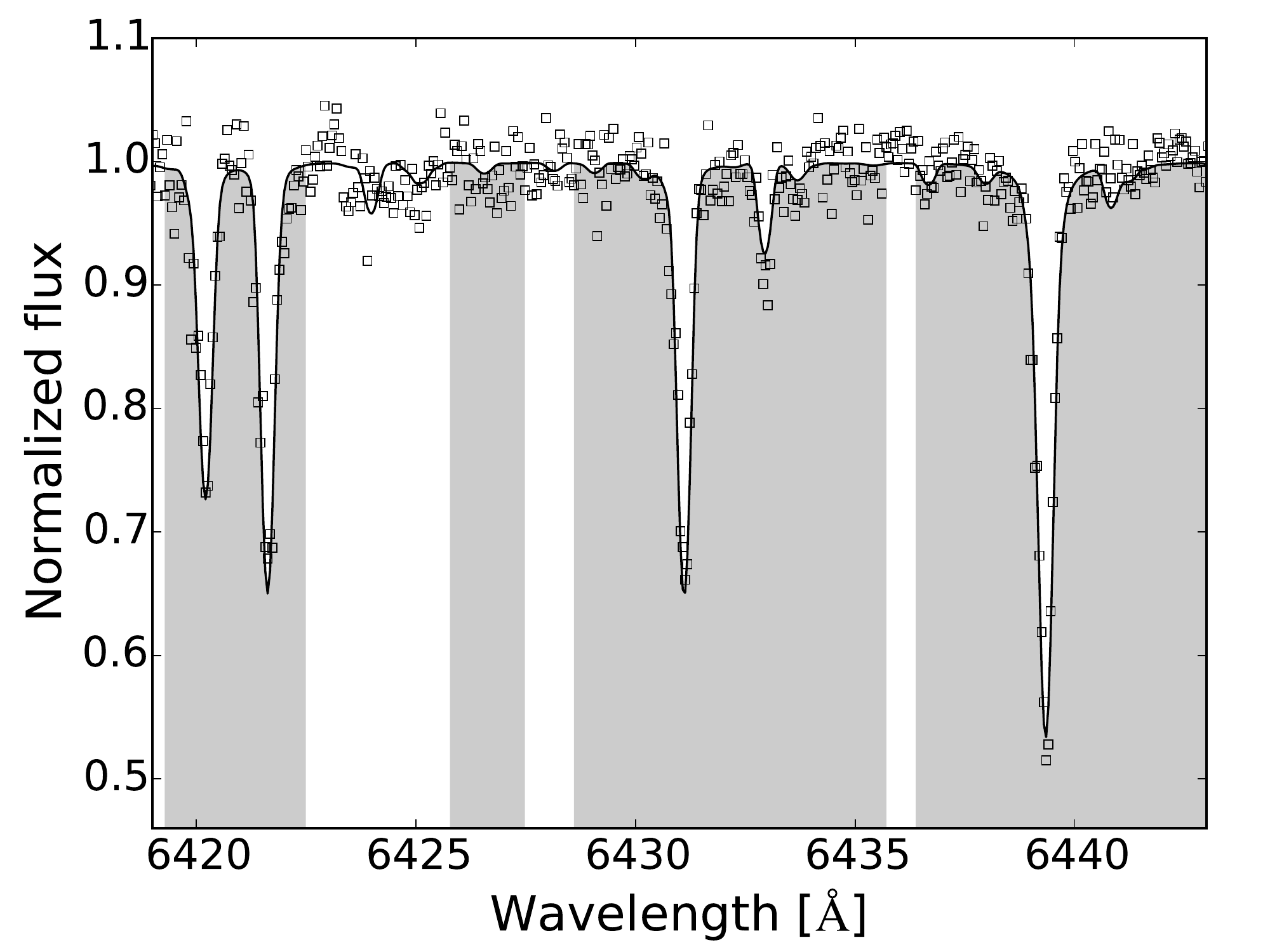}
\caption{Segment of the observed (open squares) and synthetic spectrum (solid 
line) of CoRoT 102656730. The grey shaded regions were used to determine the
stellar parameters.}
\label{figure:sme_example} 
\end{center}
\end{figure}

In our fits, we used the values for $T_{\mathrm{eff}}$, log\,$g$, [Fe/H], and
$\xi_{\mathrm{mic}}$ determined with MOOG as initial values
for the SME fit. For the macroturbulence parameters, we assumed the solar
value of $\xi_{\mathrm{mac}}=3.57\,\mathrm{km\,s^{-1}}$.
In a first fitting step, we determined the radial velocity shift
$\varv_{\mathrm{rad}}$ for all orders. We averaged the
results in the different orders and interpreted the mean as the
best estimate of the stellar parameter and its standard deviation as a
robust error estimate.
Second, we determined the mean rotational broadening parameter,
$\varv\,\mathrm{sin}\,i$, in the same manner.
In a third step, we fitted $\xi_{\mathrm{mic}}$ and [M/H], 
and finally, fitted $T_{\mathrm{eff}}$ and log\,$g$.
During the fitting process, all parameters not currently fitted remained fixed
at their best-fit values.
The final values are summarized in Table~\ref{tab:specan}.
An example of a spectral interval of CoRoT 102656730 is shown together 
with the best-fit synthetic spectrum in Fig.~\ref{figure:sme_example}.

In the case of CoRoT 102743567 we fitted $\varv_{\mathrm{rad}}$ and
$\varv\,\mathrm{sin}\,i$ using only orders covering the strong 
H$\alpha$ Balmer line and the $\ion{Na}{i}$ doublet at $\approx5990\,\AA$. 
As initial values we calculated
$T_{\mathrm{eff}}$ photometrically using B--V colors as described in
Sect.~\ref{Color, extinction, and distance}.
Furthermore, we chose typical values of main-sequence stars for log\,$g=4.3$ 
based on \citet{Gray2005}, solar metallicity, and
$\xi_{\mathrm{mic}}=1.5$\,km\,s$^{-1}$.

\subsection{Stellar parameters obtained using MOOG and SME}

In general, we find good agreement between the effective temperatures,
surface gravities, metallicities, and microturbulence velocities
derived with SME and MOOG. With a few exceptions, the values agree
within their respective uncertainties. However, the surface gravities found with
SME are systematically higher than those determined with MOOG. In particular,
we find differences between 0.12 and 0.37 for all stars but CoRoT 102778303 for which
the deviation reaches 0.88; however, the error is also large. This star is the
coolest in our sample, and we speculate that the line EW measurements may be
affected by line blends, which could also account for the rather low metallicity
of $-0.54\pm0.05$ found in our MOOG analysis. The direct fitting approach
implemented in SME should minimize errors caused by blending effects or
continuum determination.

\citet{Valenti2005} analyzed 1040 F, G,
and K stars using SME and obtained statistical errors of 44\,K on
$T_{\mathrm{eff}}$, 0.06 on log\,$g$, and 0.03 on the metallicity.
In our analysis, we found that the deviation among fits to individual echelle orders
were larger than their estimates. Thus, our SME errors are likely dominated by
systematics.

All stars show substantial photometric variability, which we attribute to
starspots. Therefore, high starspot coverage fractions are
conceivable \citep[e.g.,][]{Huber2010}, although we caution that the
data have not been taken simultaneously. A high starspot coverage fraction may
actually challenge the assumption of a single-temperature photosphere, which
tacitly underlies our spectral analysis. It may also explain at least a fraction
of the differences between the MOOG and SME analysis.

\subsection{Color, extinction, and distance}
\label{Color, extinction, and distance}

\begin{table}
\begin{center}
\caption{Intrinsic color, color excess, and photometric distance estimate}
\label{table:color_temp}
\begin{tabular}{cccc}
\hline\hline
CoRoT-ID  & (B--V)$_0$ & E(B--V) & Dist.\tablefootmark{a}
\\
 & [mag] & [mag] & [pc] \\
\hline
 102577568 & 0.70 & 0.09 & 199 \\
 102601465 & 0.69 & 0.09 & 407 \\
 102606401 & 0.53 & 0.11 & 347 \\
 102656730 & 0.62 & 0.12 & 404 \\
 102743567 & -- & -- & 474 \\
 102763571 & 0.80 & 0.05 & 291 \\
 102778303 & 0.96 & 0.18 & 138 \\
 102791435 & 0.84 & 0.12 & 207 \\
\hline
\end{tabular}
\end{center}
\tablefoot{\tablefoottext{a}{The accuracy is typically 25\,\%, see text.}}
\end{table}

\citet{Ramirez2005} provide metallicity-dependent relations between color and
effective temperature. Given the spectroscopically determined parameters and
the observed B--V colors (see Table~\ref{table:1}), we can determine the B--V
color excess. The resulting intrinsic colors, (B--V)$_0$, and excesses, E(B--V), are
given in Table~\ref{table:color_temp}. In the conversion, we applied the
effective temperatures and metallicities derived using MOOG
(Table~\ref{tab:specan}) and assumed main-sequence stars.
By substituting the upper and lower error boundaries on the
effective temperature, we estimated that the temperature-induced error on the
color excess is on the order of 0.03\,mag.

For the star CoRoT 102743567, for which no spectroscopic parameters could be determined,
we used the observed B--V color of 0.39\,mag to obtain an estimate of 7000\,K
for its effective temperature, which points to an early F-type star. We note
that \mbox{B--V$=0.39$\,mag} requires a slight stretch of the \citet{Ramirez2005}
calibration, which is formally only valid until 0.4.
This result is consistent with the classification as an A9III type star by
\citet{Sebastian2012} based on low-resolution spectroscopy.

We proceeded by converting the color excess into optical extinction by
multiplying with $R$, for which we assumed a value of 3.1 \citep{Predehl1995}.
Using Table~15.7 from \citet{allen2000}, we estimated absolute visual magnitudes
based on our effective temperature determinations and, finally,
calculated a photometric distance estimate for our target stars,
taking the extinction into account. Estimating an accuracy of $\pm 0.5$\,mag for
the distance modulus, the typical error on the distances is 25\,\%.

\subsection{Age of the sample stars}
\begin{figure}
\begin{center}
\includegraphics[width=0.49\textwidth]{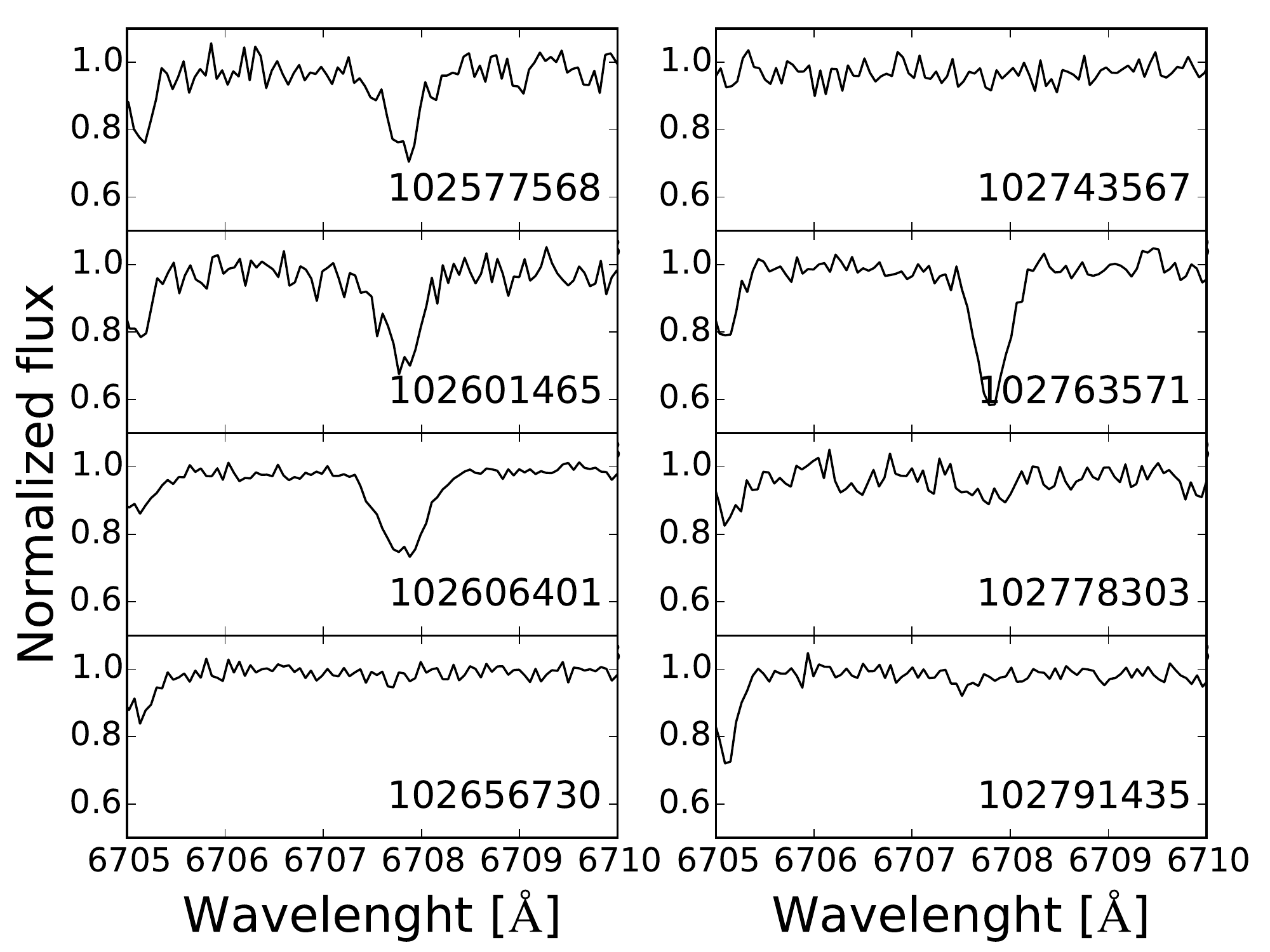}
\caption{\label{figure:lithiumlines} Normalized spectra of our target stars
showing the wavelength region around the \ion{Li}{i} line at 6708\,\AA.}
\end{center}
\end{figure}

The $\ion{Li}{i}$ resonance doublet at $6707.76\,\AA$ and $6707.91\,\AA$ can be
used as an age estimator in young stars because lithium depletion progresses
quickly during the first few hundred Myr \citep[e.g.,][]{Soderblom2010}.
In Fig.~\ref{figure:lithiumlines}, 
we show the spectral region around the $\ion{Li}{i}$ line for our sample stars.
Unambiguous detections are present for four out of the eight targets.
In a quantitative analysis, we fitted the line profile by a Gaussian
after adjusting the local continuum, which is the main source of error.

Clear detections of the lithium line were obtained in CoRoT 102577568, 
CoRoT 102601465, CoRoT 102606401, and CoRoT 102763571.
A formally significant line detection is also obtained in CoRoT 102778303. We caution,
however, that this result may require confirmation at higher S/N.
For CoRoT 102656730 and CoRoT 102791435 we derived upper limits on the line EW.
To this end, we generated artificial data sets between $6707.0\,\AA$
and $6708.5\,\AA$ assuming that no line is present and fitted them
using the model including the absorption line. Repeating this experiment
$10\,000$ times, we determined the distribution of line EW measurements given
that in reality no line exists and give the 90\,\% cut-off as the upper limit
for a significant detection.
The final EWs along with their $90\,\%$ confidence intervals are
listed in Table~\ref{table:ew}.

The $\ion{Li}{i}$ line is known to be blended by an
$\ion{Fe}{i}$ line at $6707.43\,\AA$, which has not been taken into account in
our fitting. \citet{Favata1993} studied the contribution of this iron line to the
overall EW. From their Fig.~1, we estimate that the contribution for a star with
an effective temperature of 5500\,K should be on the order of $10\,m\AA$ and
it decreases toward higher temperatures. However, for CoRoT 102778303, at a temperature of
about $4700$\,K, the $\ion{Fe}{i}$ line could have an EW of about $20$\,m\AA,
which is on the same order as our measurement. While this could indicate that
the line may indeed be attributable to iron, this seems unlikely given the
subsolar abundance pattern. All other detections
should, if anything, be affected on the $10$\,\% level, casting no doubt on the
detection of the $\ion{Li}{i}$ line itself.

\begin{figure}
\begin{center}
\includegraphics[width=0.49\textwidth]{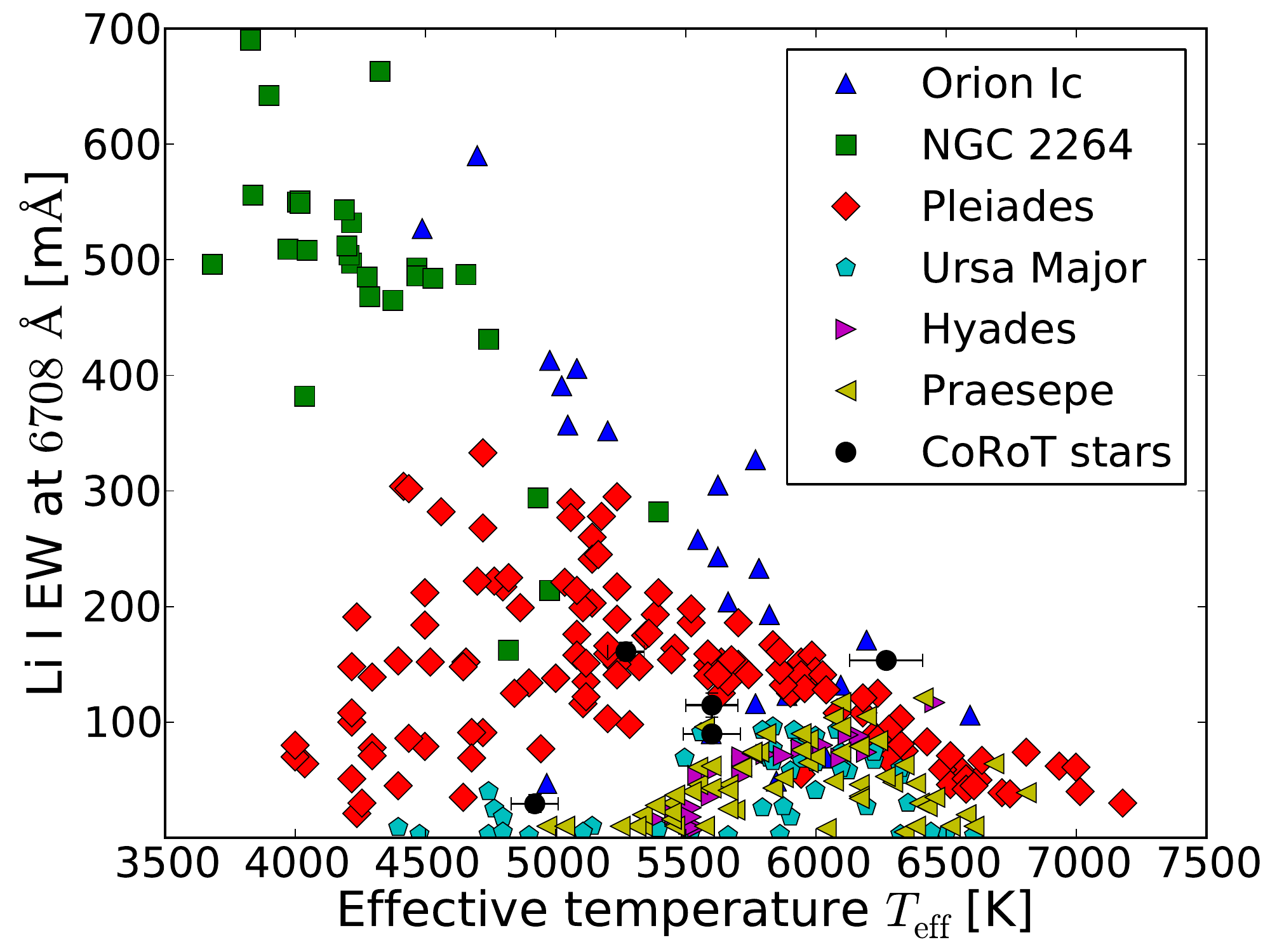}
\caption{\label{figure:age} $\ion{Li}{i}$ EW vs. effective temperature for
several open clusters. The location of our sample stars is indicated by black
circles (effective temperatures derived with MOOG).}
\end{center}
\end{figure}

We proceeded by comparing the measured $\ion{Li}{i}$ line EWs with measurements
in open cluster members of well-known age.
In Fig.~\ref{figure:age} we show our $\ion{Li}{i}$ EW measurements as a function
of effective temperature along with results for the open clusters
\object{Orion} Ic \citep[10\,Myr,][]{King1993}, \object{NGC 2264} 
\citep[10\,Myr,][]{Soderblom1999}, \object{Pleiades}
\citep[100\,Myr,][]{Soderblom1993a}, \object{Ursa Major} \citep[300\,Myr,][]{Soderblom1993b},
\object{Hyades} \citep[660\,Myr,][]{Soderblom1990}, and \object{Praesepe}
\citep[660\,Myr,][]{Soderblom1993c}. Based on their location in the plot, we arrived
at the age estimates provided in Table~\ref{table:ew}. 
Although the stars within an open cluster are formed simultaneously, there is a 
considerable scatter around a mean value in the measured EWs for a given 
effective temperature. 

\begin{table}
\begin{center}
\caption{Equivalent widths of \ion{Li}{i} line and stellar age estimates.}             
\label{table:ew}      
\begin{tabular}{cccc} 
\hline\hline          
CoRoT-ID & Li\,I EW [m\AA] & Age\tablefootmark{a} [Myr] & Age\tablefootmark{b}
[Myr] \\
\hline
102577568 & $89.8^{+6.4}_{-6.4}$    & 100--660 & $178\pm21$ \\
102601465 & $114.7^{+10.6}_{-10.3}$ & 100--660 & $234\pm29$ \\
102606401 & $153.4^{+6.5}_{-6.4}$   & 10--100 & $149\pm24$ \\
102656730 & $\le3.9$                & 300--660 & $247\pm33$ \\
102743567 & -- & -- & -- \\
102763571 & $161.0^{+8.2}_{-8.1}$   & 100 & $112\pm12$ \\
102778303 & $29.4^{+8.0}_{-7.5}$    & 300--660 & $132\pm13$ \\
102791435 & $\le3.8$                & 300--660 & $419\pm50$ \\
\hline
\end{tabular}
\end{center}
\tablefoot{\tablefoottext{a}{Comparison with open clusters};
\tablefoottext{b}{Gyrochronological age}}
\end{table}

As a result of the well-established age-activity-rotation paradigm
\citep[e.g.,][]{Skumanich1972}, rotation itself can be used as an
age indicator. We calculated the stellar ages of our target stars
based on gyrochronological models presented by \citet{Barnes2007}
(their Eq.~3). In particular, we used the rotation period and the intrinsic 
color \mbox{(B--V)$_0$} as input parameters.
The errors were estimated using their Eq.~16. Again, we present our results in
Table~\ref{table:ew}.

Almost all of the gyrochronological estimates are consistent with our estimates
based on the comparison with open clusters. The single exception is CoRoT 102778303 for
which our cluster comparison yields a higher age. However, the star CoRoT 102778303 is the
coolest target in our sample, for which we obtained effective temperature
estimates of $4840\pm100$\,K with MOOG and $4680\pm150$\,K with SME. If the star
is at the lower edge of the indicated temperature range and the \ion{Li}{i}
line is not heavily contaminated by iron, its age could also be compatible with
that of the Pleiades in Fig.~\ref{figure:age}.

\section{Discussion}
\label{section:discussion}

\subsection{Stellar parameters and age determination}
The results of our spectral analysis are broadly consistent with the
classification provided by \textit{Exo-Dat} (see Table~\ref{table:1}).

All stars show gyrochronological ages between $111$ and $418$\,Myr. Additionally,
five stars show a \ion{Li}{i} line supporting a low age, although the detection
in CoRoT 102778303 remains somewhat ambiguous.
Fast rotation and young age are compatible with high levels of activity and
large starspot coverage fractions responsible for the photometric variability.
Judging from the derived spectroscopic parameters, our sample consists of main-sequence
stars. 
However, we find that the luminosity class for CoRoT 102778303 
provided by \textit{Exo-Dat} is probably inappropriate. The obtained log\,$g$ and 
$\varv\,\sin\,i$ values, gyrochronology, and rotation period indicate that
CoRoT 102778303 should be classified as a dwarf instead of a subgiant.

\subsection{Differential rotation, spot evolution, and flip-flop events}

\begin{figure}
\begin{center}
\includegraphics[width=0.49\textwidth]{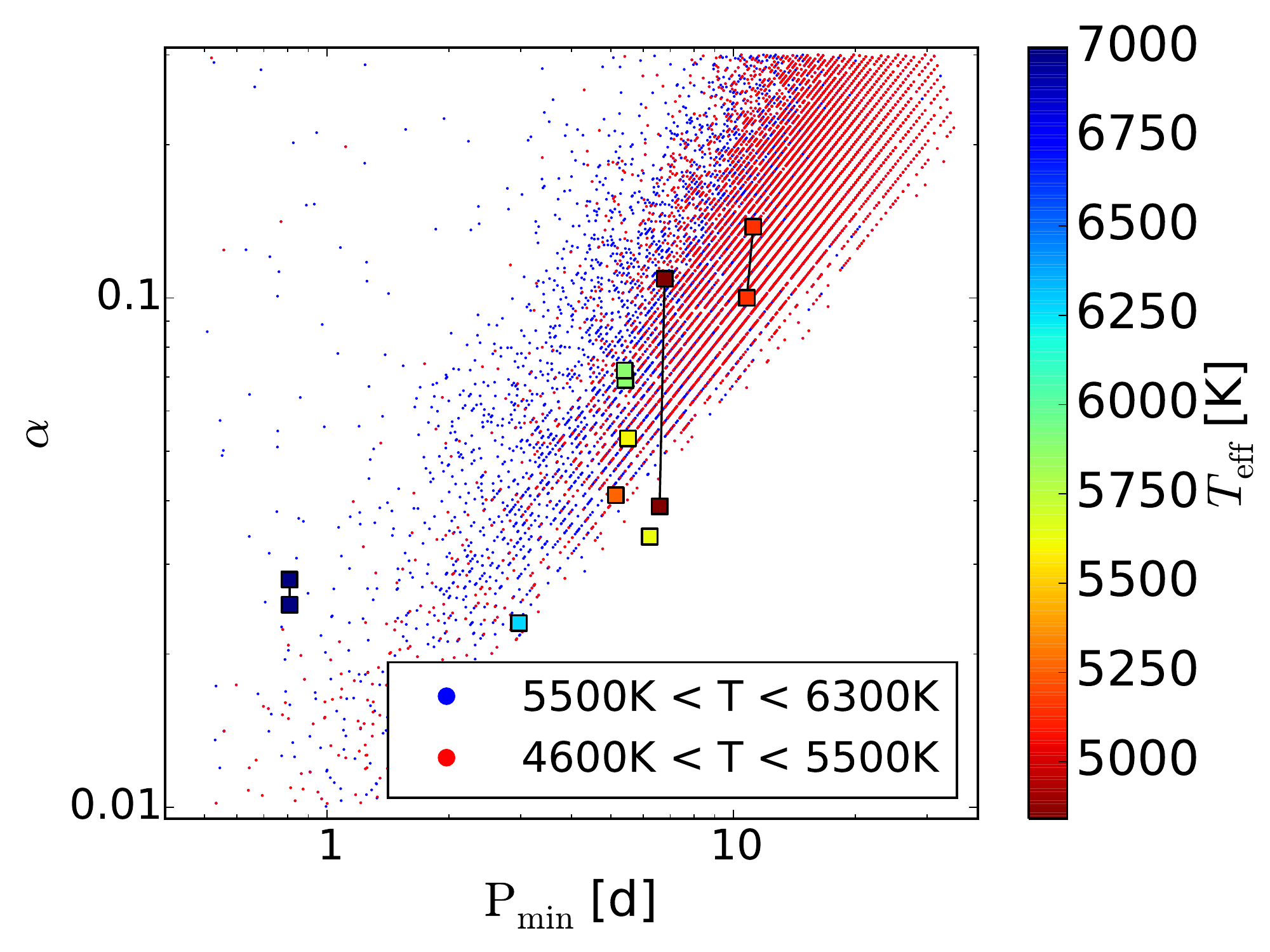}
\caption{Lower limits of the relative horizontal shear as a function
of $P_{\mathrm{min}}$ for our target stars and the sample of
\citet{Reinhold2013}. Values determined for stars with two CoRoT light
curves obtained during different epochs are connected by a solid line.
We adopted the effective temperatures derived
with MOOG and used 7000\,K for CoRoT 102743567.}
\label{figure:reinhold}
\end{center}
\end{figure}

\citet{Reinhold2013} studied differential rotation in a sample of $40\,661$
active stars observed by \textit{Kepler}. In this sample, $77.2$\,\% of the targets show a
second periodogram peak, which they attribute to a second rotation period.
Following their interpretation, we show our results for the lower limit of the
relative horizontal shear in the context of the sample presented by  
\citet{Reinhold2013} in Fig.~\ref{figure:reinhold}. Our values are clearly
compatible with theirs, although CoRoT 102601465 lies near their detection limit.
In particular, we were able to confirm the trend of increasing relative shear,
$\alpha$, with higher rotation periods. 
CoRoT 102778303 and CoRoT 102791435, 
the coolest targets with the longest rotation period which have been observed 
twice, show a considerable shift in the value of the relative shear parameter 
between the two observation epochs roughly four years apart. We
attribute the difference to both
% Beside the fact that $\alpha$ is affected by
an uncertainty resulting from the prewhitening technique and
the variability of the spot configuration. The results reflect the
uncertainty that is expected in individual measurements.

The star CoRoT 102743567 is the earliest star and fastest rotator in our sample.
Combining a canonical radius of $1.5$\,R$_{\odot}$
for the star \citep{allen2000} with the rotation period, we estimate an
equatorial rotation velocity, $\varv_{\mathrm{eq}}$, of $90-100$\,km\,s$^{-1}$.
Based on our period analysis, we further estimated a relative horizontal
shear parameter of $0.052$.
\citet{Ammler2012} studied differential rotation specifically in A- and F-type
stars using line profile analyses. Although the authors find that the fraction of
differential rotators decreases both as a function of increasing temperature
and rotation velocity, about $10$\,\% of their sample stars with $\varv\,\sin\,i\approx
100$\,km\,s$^{-1}$ show measurable differential rotation. Among these stars,
a relative horizontal shear of $\approx 0.05$ seems too moderate
(see their Fig.~10). Indeed, at $7000$\,K an absolute
horizontal shear, $\Delta \Omega$, of $0.6$\,rad\,d$^{-1}$ may be 
expected, which is also quite compatible with our value of 
$0.4$\,rad\,d$^{-1}$. We note that the spectral type of CoRoT 102743567 is in
the range of $\gamma$~Dor-type variables, which show pulsations with periods on
the order of one day \citep{Kaye1999, Zwintz2013}. Although it
is still hard to unambiguously distinguish between the rotational
modulation and a potential pulsational component \citep[cf.,][]{Zwintz2013},
our results are consistent with dominant rotational variation.

If the beating pattern is to be caused by differentially rotating active
regions (alone), its length also defines a minimal lifetime for the associated
regions, which in our case ranges between $15$\,d and about $150$\,d.
Typical sunspot lifetimes are on the order of or less than one month
\citep{Solanki2003} and low- to mid-latitude spots on rapidly rotating,
young single main-sequence stars also appear to have lifetimes of about one month
\citep{Hussain2002}. Therefore, the longer beating periods seem
challenging in terms of spot lifetimes. However, larger spots tend to have
longer lifetimes \citep{Solanki2003}.

The light curves of our sample stars (Figs.~\ref{figure:lc1} -- \ref{figure:lc3})
show a similar pattern of variability to the light curves of flip-flop stars
such as the active giant \mbox{FK Com} or young solar analogs
\citep[see][]{Jetsu1993, Berdyugina2005, Olah2006, Hackman2013}.
Flip-flops are a specific manifestation of spot evolution,
characterized by alternating spot coverage on two long-lived active longitudes
on opposing hemispheres, which requires ``coordinated''
spot evolution.
While the origin of the phenomenon may be entirely explained by the evolution of
quasi-stationary spots, differential rotation and latitudinal spot migration may
also play a role or even largely explain the observations
\citep[][]{Jetsu1993, Hackman2013}.

The surface reconstructions of \object{CoRoT-2} have revealed spot concentrations on active
longitude on opposing hemispheres \citep{Lanza2009, Huber2010}, which alternate in
strength on the beating timescale ($\approx 50$\,d). Owing to the similarity of the
light curves studied here, a comparable behavior may be expected. While the
behavior of the light curves analyzed here
is reminiscent of the flip-flop phenomenon, the flip-flop timescales
observed so far are several years \citep[e.g.,][]{Hackman2013}. For
instance, \citet{Jetsu1993} detected only three flip-flop events in
a photometric data set spanning roughly 25 years. 
While spot evolution certainly contributes to the morphology of the observed light
curves, the absence of differential rotation in our late-type sample stars appears unlikely.

\section{Summary and conclusion}
\label{section:conclusions}
 
We present a photometric and spectroscopic study of eight stars with light
curves showing photometric variability similar to that of CoRoT-2A.
The sample spans a wide range of spectral types from early F- to mid K-type.
The stellar parameters obtained from our spectral analysis with SME and MOOG
are generally consistent. For the fastest rotator in our sample, CoRoT 102743567, no
detailed spectral analysis could be carried out. For the remaining stars, we
obtained surface gravities compatible with that of main-sequence stars.
Combining the spectroscopically
derived effective temperatures with the stellar colors, we deduced distances
corrected for interstellar reddening.

The light curve analysis showed large photometric amplitudes of up to
7.5\,\% and short rotation periods between
about $0.8$\,d and $11$\,d. We found the photometric variability to be consistent
with rotational modulation by starspots.
In the majority of cases, our periodogram analysis revealed two peaks
corresponding to similar periods, whose spacing is related to the
beating period. Attributing the two periods to differentially rotating spots 
and combining the results with our
spectroscopic measurements, we find results consistent with similar previous
analyses of Kepler light curves \citep{Reinhold2013}. However, in the end it remains
unclear whether the prominent pattern of variability exhibited by the light
curves is dominated by differential rotation or spot evolution. In analogy to
findings from photometric campaigns of the active giant FK~Com, we expect both
effects to play a role.

Gyrochronological models show that all sample stars in our sample
are young dwarfs ($100-400$\,Myr). In four stars, we also found
detectable \ion{Li}{i} absorption, which also points toward a low age. This is
consistent with the high level of activity evident in the light curves.
Our sample shows a wide spread in spectral types including F-,
G-, and K-type stars, which all show a similar photometric beating behavior.
This suggests that all low-mass stars with outer 
convection zones may produce a similar CoRoT-2A-like light curve
sometime in their early evolution.

\begin{acknowledgements}
The authors thank Dr. Sebastian Schr\"oter for valuable discussions in preparing
the project and support in obtaining the spectra. This work was prepared using
PyAstronomy.
This research has also made use of the ExoDat Database, operated at LAM-OAMP, 
Marseille, France, on behalf of the CoRoT/Exoplanet program. We acknowledge 
use of observational data obtained with SARG at the TNG, Roque de Los Muchachos,
Spain. This work has made use of the VALD database, operated at Uppsala University,
the Institute of Astronomy RAS in Moscow, and the University of Vienna. Special 
thanks to Eric J. Bubar (University of Rochester) for making his MSPAWN and 
PYSPEC codes available to us. We are grateful to Nikolai Piskunov 
(University of Uppsala) for providing SME to us. We made use of the stellar 
spectrum synthesis program SPECTRUM of Richard O. Gray (Appalachian State 
University).
\end{acknowledgements}

\bibliographystyle{aa}
\bibliography{bibs}
\end{document}